\documentclass[aps,pra,twocolumn,amsmath,amssymb,superscriptaddress,floatfix]{revtex4-1}
\usepackage{amsfonts}
\usepackage[T1]{fontenc}
\usepackage{amsmath,amsbsy,amssymb}
\usepackage{times}
\usepackage{amsmath}
\usepackage{amssymb}
\usepackage[dvipdfmx]{graphicx}
\usepackage{color}

\let\mathbf=\boldsymbol

\def\bA{\bm{A}}
\def\bk{\bm{k}}

\def\bq{\bm{q}}

\def\bB{\bm{B}}
\def\bE{\bm{E}}

\def\bj{\bm{j}}

\def\EF{E_\mathrm{F}}

\def\kF{k_\mathrm{F}}

\def\Ec{E_\mathrm{c}}
\def\kB{k_\mathrm{B}}

\def\muB{\mu_\mathrm{B}}

\def\LAOSTO{LaAlO$_3$/SrTiO$_3$}

\def\bF{\bm{F}}
\def\EF{E_\mathrm{F}}
\def\EFR{E_\mathrm{FR}}
\def\Ec{E_\mathrm{c}}
\def\kFa{k_\mathrm{F1}}
\def\kFb{k_\mathrm{F2}}
\def\kFc{k_\mathrm{F3}}
\def\kB{k_\mathrm{B}}

\def\DZ{\Delta_\mathrm{Z}}
\def\DSO{\Delta_\mathrm{SO}}

\usepackage{bm}
\newcommand{\diff}{d^2}
\newcommand{\imag}{\mathrm{Im}\,}
\newcommand{\real}{\mathrm{Re}\,}

\newcommand{\imu}{i}
\newcommand{\epn}{e}

\newcommand{\dg}{\dagger}
\newcommand{\la}{\langle}
\newcommand{\ra}{\rangle}
\newcommand{\al}{\alpha}
\newcommand{\sg}{\sigma}
\newcommand{\gm}{\gamma}

\newcommand{\permutation}[2]{{}_{#1}\textrm{P}_{#2}} 
\newcommand{\mrm}{\mathrm}

\begin{document}

\title{Nonreciprocal charge transport in two-dimensional 
noncentrosymmetric superconductors}

\author{Shintaro Hoshino}
\affiliation{Department of Physics, Saitama University,
Shimo-Okubo 255, Sakura-ku, Saitama 338-8570, Japan}
\author{Ryohei Wakatsuki}
\affiliation{Department of Applied Physics, University of Tokyo, Hongo 7-3-1, Bunkyo-ku, 
Tokyo 113-8656, Japan}
\author{Keita Hamamoto}
\affiliation{Department of Applied Physics, University of Tokyo, Hongo 7-3-1, Bunkyo-ku, 
Tokyo 113-8656, Japan}
\author{Naoto Nagaosa}
\affiliation{RIKEN Center for Emergent Matter Science (CEMS), Wako, Saitama 351-0198, Japan}
\affiliation{Department of Applied Physics, University of Tokyo, Hongo 7-3-1, Bunkyo-ku, 
Tokyo 113-8656, Japan}

\date{\today}

\begin{abstract}
Nonreciprocal charge transport phenomena are studied theoretically for two-dimensional noncentrosymmetric superconductors under an external magnetic field $B$. Rashba superconductors, surface superconductivity on the surface of three-dimensional topological insulators (TIs), and transition metal dichalcogenides (TMD) such as MoS$_2$ are representative systems, and the current-voltage $I$-$V$ characteristics, i.e., $V=V(I)$, for each of them is analyzed. $V(I)$ can be expanded with respect to the current $I$ as $V(I)= \sum_{j=1,\infty}  a_j(B,T) I^j$, and the $(B,T)$-dependence of $a_j$ depends on the mechanism of the charge transport. Note that the magnetochiral anisotropy in the normal state is expressed by $a_1(B,T)= R_0$ and $a_2(B,T) = R_0 \gamma B$ with the constant $\gamma$. Our analysis is based on the time-dependent Ginzburg-Landau (TDGL) theory, which contains up to third order terms in the momentum of the order parameter. Above the mean field superconducting transition temperature $T_0$, the fluctuation of the superconducting order parameter gives the additional conductivity, i.e., paraconductivity. Extending the analysis of paraconductivity to the nonlinear response, we obtain the nonreciprocal charge transport expressed by $a_2(B,T) = a_1(T) \gamma(T) B$, where $\gm$ converges to a finite value at $T=T_0$.  Below $T_0$, the motion of vortices is relevant to the voltage drop, and the dependence of $a_1$ and $a_2$ on $B$ and $T$ is different depending on the system and mechanisms. For Rashba superconductors and superconducting surface state of three dimensional TIs under the in-plane magnetic field, the Kosterlitz-Thouless (KT) transition occurs at $T_{\rm KT}$, below which the vortices and anti-vortices are bound and the resistivity becomes zero. Therefore, $a_1(B,T)$ and $a_2(B,T)$ are defined only above $T_{\rm KT}$. In this case, the vortex contributions to the transport coefficients are $a_1 (B,T) = {\rm const.}$ and $a_2(B,T) \sim B (T_0-T)^{-1} $ for $T\rightarrow T_0$. It is also found near $T_{\rm KT}$ that $a_1(B,T) \sim \exp[ - (b\tau_c/\tau)^{1/2} ]$ and $a_2(B,T) \sim B \tau^{-3/2} \exp[ - (b\tau_c/\tau)^{1/2} ]$ with the reduced temperatures $\tau(T) = (T-T_{\rm KT})/T_{\rm KT}$, $\tau_c = \tau(T_0)$, and an order of unity constant $b$. Below $T_{\rm KT}$, both $a_1(B,T)$ and $a_2(B,T)$ vanish, and $a_3(B,T) \sim {\rm const.}$ and $a_4(B,T) \sim B$ become the leading two terms in the expansion. On the other hand, for TMD with the magnetic field perpendicular to the 2D system, the  KT transition is gone and the system remains resistive even well below $T_{\rm KT}$. In this case, there are two possible mechanisms for the nonreciprocal charge transport. One is the anisotropy of the damping constant for the motion of the vortex induced by the external magnetic field and current. In this case, $a_1(B,T) \sim B$ and $a_2(B,T) \sim B^2$. The other one is the ratchet potential acting on the vortex motion, which gives $a_1(B,T) \sim B$ and $a_2(B,T) \sim B$. Based on these results, we propose the experiments to identify the mechanism of the nonreciprocal transport with the realistic estimates for the order of magnitude of the coefficients $a_1(B,T)$ and $a_2(B,T)$ for each case. 
\end{abstract}

\maketitle

\section{Introduction} \label{Sec:Introduction}
The nonreciprocal charge transport in noncentrosymmetric systems is a fundamental
and important issue. It is deeply related to the broken symmetry of 
spatial inversion $P$ and the time-reversal $T$.
For the linear response, the microscopic time-reversal symmetry
leads to the Onsager's reciprocal theorem \cite{Onsager,Kubo} given by
\begin{equation}
    K_{A B} (q, \omega, B) = \epsilon_A \epsilon_B K_{B A} (-q, \omega, -B),
\label{Eq:Onsager}
\end{equation}
where $K_{AB}(q, \omega,B)$ describes the linear response of the physical observable $A$
to the field coupled to the observable $B$ with the wavevector $q$ and frequency $\omega$
under the magnetic field $B$ (which breaks the time-reversal symmetry).
$\epsilon_{A}= \pm 1$ ($\epsilon_{B}$ ) specifies the even (1) or odd (-1) nature of the
observable $A$  ($B$) with respect to $T$.
On the other hand, the spatial inversion symmetry gives 
\begin{equation}
    K_{A B} (q, \omega, B) = \eta_A \eta_B K_{A B} (-q, \omega, B),
\label{Eq:Inversion}
\end{equation}
with $\eta_A$ ($\eta_B$) being the analogous quantity to 
$\epsilon_{A}= \pm 1$ ($\epsilon_{B}$) for $P$.
When both of $T$ and $P$ are broken, the Onsager reciprocal theorem 
allows the directional linear response of the diagonal response.
For example, the dielectric function for light can have the form
$\varepsilon_{\mu \mu}(q, \omega, B) = \varepsilon_0 + \alpha B q$
which describes the directional dichroism of unpolarized light.

Rikken extended this consideration to the nonlinear response by heuristic argument,
i.e., replacing the wavevector $q$ by the current $I$, leading to the expression 
of the resistivity \cite{Rikken2,Morimoto}
\begin{equation}
    R = R_0 ( 1 + \gamma B I).
 \label{Eq:Rikken}
\end{equation}
The coefficient $\gamma$, which is called $\gm$-value in the following, is usually a rather small value of the order
of $\sim 10^{-2}$ to $~ 10^{-1}$T$^{-1}$A$^{-1}$ \cite{Rikken1,Krstic,Rikken3,Pop}. 
This is because
the nonreciprocal transport requires both the magnetic energy $\mu_B B$ and
the spin-orbit interaction $\lambda$, which are small compared with the
energy denominator, i.e., kinetic energy of the electrons 
(typically the Fermi energy $E_{\rm F}$).
To enhance the nonreciprocal transport, there are two ways. One is to 
reduce the energy denominator and the other is to enlarge the spin-orbit
interaction. This is realized in BiTeBr with the giant bulk Rashba
splitting by reducing the electron density \cite{IdeueHamamoto}. Furthermore,
in the superconductors, the Fermi energy is replaced by the 
energy gap in the energy denominator, which leads to the huge enhancement
of $\gamma$ as demonstrated in MoS$_2$ \cite{Wakatsuki}. The theoretical analysis, however,
is limited to the paraconductivity above the mean field transition temperature $T_0$,
and that below $T_0$ still remains an important unresolved issue although 
the experiment shows the further increasing value of $\gamma$ there \cite{Wakatsuki}.

In the present paper, we give a comprehensive and unified treatment
of the nonreciprocal charge transport in two-dimensional 
noncentrosymmetric superconductors (2DNS).
There are several possible mechanisms for it, and accordingly we generalize
Eq.(\ref{Eq:Rikken}) to the current ($I$)-voltage ($V$) characteristics as 
\begin{equation}
V(I)= 
\sum_{j=1}^{\infty}  
a_j(B,T) I^j.
 \label{Eq:IV}
\end{equation}
We take 
(i) the Rashba superconductors, and surface superconductivity on 
the surface of three-dimensional topological insulators (TIs), and
(ii) the transition metal dichalcogenides (TMD) such as MoS$_2$, 
as the two 
representative examples of 2DNS.

In 2D superconductors, there are two characteristic temperatures.
One is the mean field transition temperature $T_0$, below which the 
amplitude of the order parameter develops, and the other is the
Kosterlitz-Thouless (KT) transition temperature $T_{\rm KT}$,
below which the vortices and anti-vortices are bound and the 
resistivity becomes zero. 
The behavior of the resistivity $R(T)$ as a function of the temperature
is well described by
\begin{equation}
R(T) \cong 
2.7
R_n 
( \xi_c/\xi_{+}(T))^2
,
\end{equation}
with $R_n$ being the normal state resistivity and $\xi_+(T)$ is
the diverging coherent length 
$\xi_+(T) = \xi_c b^{-1/2} \sinh[(b\tau_c/\tau)^{1/2}]$
where the reduced temperatures are introduced by $\tau=(T-T_{\rm KT})/T_{\rm KT}$ and 
$\tau_c=(T_0-T_{\rm KT})/T_{\rm KT}$ \cite{Halperin79}.
The parameter $\xi_c$ is a coherence length obtained by the GL theory and $b$ is an order of unity constant.
With the external magnetic field $B$, the situation depends on its 
direction.
With the in-plane $B$, only the Zeeman effect is relevant, and
KT transition survives with the reduced transition temperature $T_{\rm KT}(B)$.
This is the case for (i) to obtain the nonreciprocal transport.
On the other hand, with the out-of-plane magnetic field
considered in the case (ii), the vortices 
are introduced even below $T_{\rm KT}$, and the system remains
resistive down to low temperatures. 
The experiment for MoS$_2$ \cite{Wakatsuki} employs this configuration, and
continue to define $\gamma$ in Eq.~\eqref{Eq:Rikken}, although
the $B$-dependence of $a_2(B,T)$ is different for different mechanisms
as shown below.

\begin{table*}[t]
\begin{center}
\includegraphics[width=160mm]{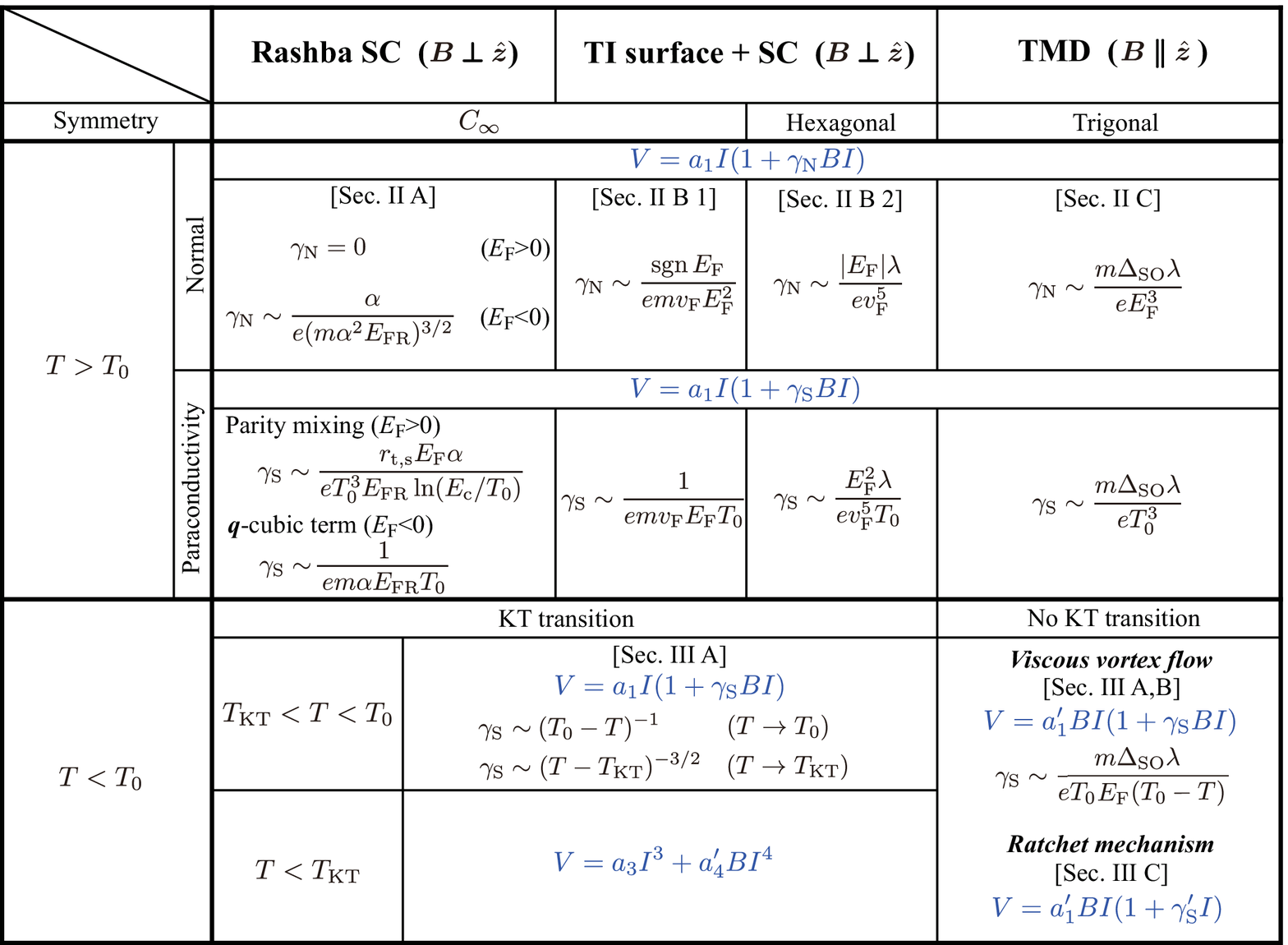}
\caption{
Summary of nonreciprocal charge transport in two-dimensional noncentrosymmetric superconductors at low magnetic field $B$.
The parameters are electron charge $e$ and band mass $m$, Fermi energy $E_{\rm F}$, Fermi velocity $v_{\rm F}$, Rashba parameter $\al$ (we also define $E_{\rm FR} = 2E_{\rm F} + m \al^2$), 
triplet or singlet pair mixing parameter $r_{\rm t,s}$, cutoff energy $E_{\rm c}$, 
hexagonal or trigonal warping parameter $\lambda$, spin-orbit splitting $\Delta_{\rm SO}$, mean-field superconducting transition temperature $T_0$, and Kosterlitz-Thouless transition temperature $T_{\rm KT}$.
The unit system $\hbar=k_{\rm B}=\mu_{\rm B}=1$ is used.
The sample width $W$ is omitted from the expressions.
The temperature dependence for the paraconductivity contributions at $T>T_0$ enters through the formula in Eq.~\eqref{eq:temp_dep_paracon}.
See each section for more details.
}
\label{fig:table}
\end{center}
\end{table*}

The obtained results for nonreciprocal charge transport are summarized in Table~\ref{fig:table}, which will be discussed in greater detail in the rest of this paper.
 Our plan 
is as follows. In section II, we study the paraconductivity above the
 mean field transition temperature $T_0$ in terms of the Ginzburg-Landau theory 
derived for several cases of interest. In section III we present the 
analysis of the charge transport below $T_0$, where the 
vortex motion is responsible for the voltage under the current flow. 
There are three mechanisms for the nonreciprocal response, i.e.,
(i) the change in the KT transition temperature due to the current, 
(ii) the modified dissipation for the vortex motion due to the current, and 
(iii) the Ratchet potential for the vortex.  
Sections IV and V are devoted to the 
discussion and conclusions, respectively.
The detailed derivation of GL free energy is given in Appendix A.
Appendices B and C describe the effects of impurity and Landau levels, respectively, for TMD.

\section{Ginzburg-Landau theory and paraconductivity in noncentrosymmetric 
superconductors} \label{Sec:GL}

We discuss the nonreciprocal current in the temperature regime slightly above the mean 
field critical temperature ($T \gtrsim T_0$).
In this regime, the charge current is mainly carried by the thermal fluctuation of the 
superconducting order parameter.
Such excess conductivity is called paraconductivity \cite{SkocpolTinkham, LarkinVarlamov}.
The paraconductivity consists of the Aslamazov-Larkin (AL) contribution \cite{AslamazovLarkin} 
and the Maki-Thompson (MT) contribution \cite{Maki, Thompson}.
Although fully quantum treatment is necessary for the MT term, the AL term can be 
discussed by the Ginzburg-Landau (GL) theory 
\cite{SkocpolTinkham, LarkinVarlamov, Schmid}.

The nonreciprocity of the paraconductivity has been studied in two-dimensional 
TMD \cite{Wakatsuki} and Rashba systems \cite{WakatsukiNagaosa}.
It is expected that the MT term is smaller than the AL term under a magnetic field 
due to pair breaking effect \cite{Yip90} by external field,
and the explicit forms of the $\gamma$-values in Eq.(\ref{Eq:Rikken})
have been calculated based on the GL theory.
Two different origins of the nonreciprocal current have been discussed in these 
materials, i.e., the trigonal warping of the band structure in the 
TMD \cite{Wakatsuki}, and the parity mixing of the singlet and triplet 
order parameters \cite{BauerSigrist, Yip} in the 
superconducting Rashba systems.
In both of the systems, the nonreciprocity is markedly enhanced in the superconducting 
fluctuation regime because of the scale difference between the Fermi energy and the 
superconducting gap.

In this section, we discuss the nonreciprocal paraconductivity in five models, the Rashba superconductors with and without parity mixing, the surface state of topological insulator with the parabolic dispersion and the hexagonal warping, and the two-dimensional 
TMD \cite{Wakatsuki}.
We will also compare these results with the normal contribution, to demonstrate the enhancement of nonreciprocal signal in the superconducting fluctuation regime.
While some of the results in the following are not entirely new as they have been published previously \cite{IdeueHamamoto, Wakatsuki, WakatsukiNagaosa}, here we have briefly provided them again in order to 
make this article self-contained and complete.

In the following, we derive an expression for the $\gm$-value in the region where the paraconductivity dominates over the normal conductivity near the mean-field temperature $T_0$.
With the expressions for the conductivities defined by $j = \sg_1 E + \sg_2 E^2$, the deviation from this limit can be included by replacing the fluctuation contribution as $\sg_1 \longrightarrow \sg_1 + \sg_n$ where $\sg_n$ is a normal conductivity.
Then the temperature dependence for $\gm_{\rm S} \propto \sg_2/(\sg_1+\sg_n)^2$ enters as
\begin{align}
\gm_{\rm S}(T) = \gm_{\rm S} \left[
1 + \frac{1}{c_0}\, \frac{\sg_n}{e^2/h}\, \frac{T-T_0}{T_0}
\right]^{-2}
, \label{eq:temp_dep_paracon}
\end{align}
where $c_0$ is an order-of-unity constant and $h$ is a Planck constant.
This formula is relevant for the wider temperature range.
The detailed functional forms can be found in the rest of this section.

\subsection{Rashba superconductor} \label{sec:rashba_sc}

\subsubsection{Parity mixing}

In the following of this paper, we use the unit system $\hbar = \kB = \muB = 1$ unless otherwise specified explicitly.
In Ref. \onlinecite{WakatsukiNagaosa}, the nonreciprocal current due to the parity mixing of 
the superconducting order parameter has been studied in the Rashba 
superconductors.
The normal state Hamiltonian is
\begin{equation}
H = \frac{\bk^2}{2m} + \alpha \left(k_x \sigma_y - k_y \sigma_x\right) - \bB \cdot \bm{\sigma}
\label{Eq:rashba},
\end{equation}
with $m$, $\bk$, $\alpha$, $\bB$ and $\bm \sigma$ being the electron mass, electron wavenumber, Rashba parameter, in-plane magnetic field, and the Pauli matrix for real spins, respectively.
If the Rashba splitting $m \alpha^2$ is 
larger than $T_0$ or the superconducting gap, only the pairing in each spin-split band is relevant.
Then, if both even parity and odd parity interactions exist, the parity of the superconducting 
order parameters is mixed.

To make the discussion simple, we fix the form of the interaction Hamiltonian in the band basis as \cite{WakatsukiNagaosa}
\begin{equation}
    H_\mathrm{int} = -\sum_{\bk \bk' \lambda \lambda'}
        t_{\bk \lambda} t_{\bk' \lambda'}^* \hat{g}_{\lambda \lambda'}
        \psi^\dagger_{\bk \lambda} \psi^\dagger_{-\bk \lambda}
        \psi_{-\bk' \lambda'} \psi_{\bk' \lambda'}, \label{Eq:Hint}
\end{equation}
where $\psi^\dagger_{\bk \lambda}$ and $\psi_{\bk \lambda}$ are the creation and annihilation operators with the band index $\lambda = \pm$, and $t_{\bk\lambda} = \lambda i \mathrm{e}^{i\phi_{\bk}}$ with $\phi_{\bk} = \mathrm{arg} (k_x+i k_y)$.
We have assumed that the spin-splitting due to the Rashba interaction is much larger than the superconducting mean-field temperature ($m\al^2 \gg T_0$), and hence, the inter-band pairings are neglected.

We have assumed the interactions in the spin-basis which makes the matrix $\hat{g}$ independent on $\bk$.
The even-parity channel is the standard BCS type on-site interaction,
\begin{equation}
    -V^\mathrm{g} \sum_{\bk \bk'} c^\dagger_{\bk \uparrow}c^\dagger_{-\bk \downarrow} c_{-\bk' \downarrow}c_{\bk' \uparrow},
\end{equation}
where $c^\dagger_{\bk \sigma}$ and $c_{\bk \sigma}$ are the creation and annihilation operators of the electron with momentum $\bk$ and spin $\sigma$.
The odd-parity channel is
\begin{equation}
    -\sum_{\bk\bk'}V_{ij}^\mathrm{u}\left(\bk, \bk'\right)\left(i\sigma_i\sigma_2\right)_{\alpha\beta}\left(i\sigma_j\sigma_2\right)_{\gamma\delta} c^\dagger_{\bk \alpha}c^\dagger_{-\bk \beta} c_{-\bk' \gamma}c_{\bk' \delta},
\end{equation}
with $V^\mathrm{u}_{ij}\left(\bk, \bk'\right) = V^\mathrm{u}\hat{\gamma}_i\left(\bk\right) \hat{\gamma}_j\left(\bk'\right)$ and $\hat{\bm \gamma}\left(\bk\right) = \frac{1}{k}\left(-k_y, k_x\right)$.
Then, the matrix $\hat{g}$ in Eq. (\ref{Eq:Hint}) is
\begin{equation}
    \hat{g} = 
    \begin{pmatrix}
        g_1 & g_2 \\
        g_2 & g_1
    \end{pmatrix},
\end{equation}
with  $g_1 = \left(V^\mathrm{g}+V^\mathrm{u}\right)/4$ $(> 0)$ and $g_2 = \left(V^\mathrm{g}-V^\mathrm{u}\right)/4$.

In the following, we focus on two limiting cases.
(1) $|V^\mathrm{u}| \ll |V^\mathrm{g}|$ (or equivalently, $g_2 \approx g_1$) case. The singlet pairing is dominant and the triplet mixing is proportional to the parameter $r_\mathrm{t} = \frac{g_1-g_2}{g_1}$, which we treat perturbatively.
(2) $|V^\mathrm{u}| \gg |V^\mathrm{g}|$ (or equivalently, $g_2 \approx -g_1$) case. In this case, the triplet pairing is dominant and the singlet mixing parameter is given by $r_\mathrm{s} = \frac{g_1+g_2}{g_1}$.

In order to treat the parity mixing, the two-component GL theory has been employed.
We assume that the Fermi energy $E_{\rm F}$ is on the conduction band ($E_{\rm F}>0$) since the nonreciprocal paraconductivity vanishes for $E_{\rm F} <0$ \cite{WakatsukiNagaosa} 
in sharp contrast to that in the normal state \cite{IdeueHamamoto}.
The derivation of the GL free energy is shown in Ref. \onlinecite{WakatsukiNagaosa}, and the result is
\begin{equation}
    F = \int\frac{d^2\bq}{\left(2\pi\right)^2}\sum_{\lambda \lambda'} \Psi^*_\lambda 
    \left[ \left(\hat{g}^{-1}\right)_{\lambda \lambda'} + \delta_{\lambda\lambda'}N_{\lambda} 
\left(S_1 - L_{\lambda\bq}\right) \right] \Psi_{\lambda'}, \label{Eq:Frashba1}
\end{equation}
with the parameters
\begin{align}
    N_\lambda &= \frac{m}{2\pi}\left(1-\lambda\frac{\sqrt{m\alpha^2}}{\sqrt{\EFR}}\right), \\
    L_{\lambda\bq} &= K_\lambda \bq^2 - \lambda R_\lambda \left(B_y q_x - B_x q_y\right), \\
    K_- &= K_+ = \frac{S_3 \EFR}{8m}, \\
    R_- &= R_+ = \frac{S_3 \sqrt{\EFR}}{2\sqrt{m}}, \\
    S_1 &= \log \frac{2\mathrm{e}^{\gamma_\mathrm{E}} \Ec}{\pi T}, \\
    S_3 &= \frac{7\zeta\left(3\right)}{4\left(\pi T\right)^2}.
\end{align}
We have defined
$\EFR = 2\EF+m\alpha^2$, and $\gamma_\mathrm{E}$ is the Euler's constant, and $\Ec$ is the cutoff energy.

Then, the paraconductivity can be calculated as \cite{Schmid, Wakatsuki, WakatsukiNagaosa}
\begin{align}
\bj &= -T \sum_{\bq}C \left.\frac{\partial f\left(\bq+2e\bA\right)}{\partial \bA}
\right|_{\bA = \bm{0}} \notag \\
& \times \int_{-\infty}^0 du \exp \left[-C \int_u^0 dt f \left(\bq - 2e\bE t\right)\right] 
, \label{Eq:paracon}
\end{align}
with $C = \frac{32T_0}{\pi \nu}$ and the density of states $\nu$ at the Fermi level.
The function $f$ is the eigenvalue of the matrix in Eq. (\ref{Eq:Frashba1}) with a higher critical temperature.
In the case of a single component GL free energy, $f$ is defined as $F = \int \frac{d^2\bq}{\left(2\pi\right)^2} f |\Psi|^2$.
The paraconductivity is shown to be
\begin{align}
    j_x &= \sigma_1 E_x + \sigma_2 E_x^2, \label{Eq:j} \\
    \sigma_1 &= \frac{e^2}{16\epsilon}, \label{Eq:j1}\\
    \sigma_2 &= \frac{\pi e^3 B_y r_{\mathrm{t,s}}}{128\epsilon^2} \notag \\
    &\times
    \frac{N_-N_+\left(K_-N_--K_+N_+\right)\left(K_-R_++K_+R_-\right)}{S_1\left(T_0\right)T_0\left(N_-+N_+\right)\left(K_-N_-+K_+N_+\right)^2}, \label{Eq:j2}
\end{align}
in the lowest order of $r_\mathrm{t,s}$, and $\epsilon=\frac{T-T_0}{T_0}$ is the reduced temperature.
It should be noted that the sign of the nonreciprocal current depends on the sign of $r_\mathrm{t,s}$, i.e., sign of $V^\mathrm{u}$ in the case of singlet dominant case and $V^\mathrm{g}$ in the case of triplet dominant case.
The $\gamma$-value (see Eq. (\ref{Eq:Rikken})) is obtained by 
$W \gamma_{\rm S} = \sigma_2 / \left(B \sigma_1^2\right)$ with $W$ being the sample width \cite{IdeueHamamoto, Wakatsuki, WakatsukiNagaosa}.
The explicit form is
\begin{equation}
    W \gamma_\mathrm{S} = \frac{\pi r_\mathrm{t,s} S_3 \EF \alpha}{e S_1T_0 \EFR}. 
\label{Eq:gammapm}
\end{equation}

For normal state, on the other hand, it can be shown that the nonreciprocal current exists in $\EF < 0$
within the Boltzmann theory,
where the $\gm$-value is given by $W\gm_{\rm N}\sim |\al| / e(m\al^2 E_{\rm FR})^{3/2}$ \cite{IdeueHamamoto}.
In the case of $\EF > 0$, the normal contribution is zero and there is only the paraconductivity contribution for the superconducting fluctuation regime.
If we assume that the Rashba splitting is comparable to the Fermi energy in both cases, the ratio 
is $\gamma_\mathrm{S}(E_{\rm F}>0) / \gamma_\mathrm{N}(E_{\rm F}<0) \sim 
r_{\rm t,s}
|\EF|^3 / \left(S_1 T_0^3 \right)$.

We emphasize that the above nonreciprocity originates from the parity mixing ($r$ in Eq. 
(\ref{Eq:gammapm})).
In the next subsection, we will show another mechanism, which relies on the cubic term with respect to the momentum of the superconducting order parameter.

\subsubsection{
$\bm q$-cubic term
}
Now we consider the nonreciprocal paraconductivity in the Rashba superconductor without parity mixing.
In contrast to the previous subsection, we consider the case only for the $s$-wave pairing.
Although the expansion up to the second order of the momentum of the order parameter does 
not create finite nonreciprocal current, we have finite nonreciprocal current by considering the expansion up 
to the third order.
The GL free energy for the $s$-wave order parameter can be obtained from the two-component 
GL free energy written with the band basis.
If we diagonalize the two-component GL free energy in the band basis, the free energy for the 
$s$-wave order parameter is found to be a sum of the diagonal components.

We first define the three Fermi wavenumbers and the density of states which correspond to (1) the upper band, (2) inner branch of the lower band, and (3) outer branch of the lower band:
\begin{align}
\kFa &= -m\alpha + \sqrt{m \EFR}, \\
\kFb &= m\alpha-\sqrt{m \EFR}, \\
\kFc &= m\alpha+\sqrt{m \EFR}, \\
\nu_1 &= \frac{m}{2\pi}\left(1-\alpha\sqrt{m/\EFR}\right), \\
\nu_2 &= \frac{m}{2\pi}\left(-1+\alpha\sqrt{m/\EFR}\right), \\
\nu_3 &= \frac{m}{2\pi}\left(1+\alpha\sqrt{m/\EFR}\right).
\end{align}
Then, we can derive the contribution from each branch separately, whose example is shown in Appendix A.
To write the GL free energy in a simple form, 
we define the following functions,
\begin{align}
f_A &= \frac{\EFR}{8m}\bq^2 + \frac{3 B_y}{32m\sqrt{m\EFR}}
\left(5+\frac{3m\alpha}{-m\alpha+\sqrt{m\EFR}}\right)q_x \bq^2, \\
f_B &= \frac{\EFR}{8m}\bq^2 + \frac{3 B_y}{32m\sqrt{m\EFR}}
\left(-5+\frac{3m\alpha}{m\alpha+\sqrt{m\EFR}}\right)q_x \bq^2,
\end{align}
where we have assumed $B_x = 0$ because it does not affect the conductivity along the $x$ direction.
In the case of $\EF > 0$, the free energy is then
\begin{equation}
F = \int \frac{d^2\bq}{\left(2\pi\right)^2} \left[ \frac{1}{g} - \left(\nu_1+\nu_3\right) 
S_1 + S_3 \left(\nu_1 f_A + \nu_3 f_B\right) \right] |\Psi_{\bq}|^2,
\label{eq:free_energy_Rashba_cubic}
\end{equation}
with $g$ being the amplitude of the attractive interaction.
Although the $\bm q$-linear term in general appears, it can be absorbed by a constant shift in $\bm q$-space and does not explicitly appear in the final results.
The paraconductivity obtained from Eq. (\ref{Eq:paracon}) is given by
\begin{equation}
j_x = \frac{e^2}{16\epsilon}E_x + 
\frac{3\pi e^3 \alpha B_y}{512\EFR^2T_0\epsilon^2}E_x^2,
\end{equation}
Correspondingly, the $\gamma$-value is
\begin{equation}
W \gamma_\mathrm{S} = \frac{3\pi \alpha}{2e\EFR^2 T_0}.
\end{equation}
We also consider the case of $\EF < 0$.
The free energy is given by
\begin{equation}
F = \int \frac{d^2\bq}{\left(2\pi\right)^2} \left[ \frac{1}{g} - 
\left(\nu_2+\nu_3\right) S_1 + S_3 \left(\nu_2 f_A + 
\nu_3 f_B\right) \right] |\Psi_{\bq}|^2.
\end{equation}
Accordingly, we obtain
\begin{equation}
j_x = \frac{e^2}{16\epsilon}E_x + \frac{15\pi e^3 B_y}{1024 m 
\alpha \EFR T_0\epsilon^2}E_x^2,
\end{equation}
and the $\gamma$-value is
\begin{equation}
W \gamma_\mathrm{S} = \frac{15\pi}{4e m \alpha \EFR T_0}. \label{Eq:gammacubic}
\end{equation}
We have shown that the simple Rashba model Eq. (\ref{Eq:rashba}) has the nonreciprocal 
current if we consider the GL free energy up to $O(\bq^3)$.
This mechanism is different from the previous subsection \cite{WakatsukiNagaosa}, where the singlet and triplet parity mixing is essential for the nonreciprocal current.

Let us compare the nonreciprocal paraconductivities with and without parity mixing.
The ratio between the $\gamma$-value from the parity mixing mechanism Eq. (\ref{Eq:gammapm}) (denoted as $\gamma_\mathrm{S}^{\mathrm{pm}}$), and the $\gamma$-value from the cubic term Eq. (\ref{Eq:gammacubic}) (denoted as $\gamma_\mathrm{S}^{\mathrm{c}}$) is
\begin{equation}
\frac{\gamma_\mathrm{S}^{\mathrm{pm}}}{\gamma_\mathrm{S}^{\mathrm{c}}} \sim 
\frac{r_\mathrm{t,s} \EF \EFR}{S_1 T_0^2}. 
\label{Eq:pmcratio}
\end{equation}
If the even parity interaction corresponds to the on-site interaction and the odd parity interaction corresponds to the nearest-neighbor interaction, their amplitudes are roughly estimated as $e^2 / a_0$ and $e^2 / a$ with $a_0$ and $a$ being the Bohr radius and the lattice constant, respectively.
Therefore, $r_{\rm t,s} \sim 0.1$ is reasonable value for the singlet dominant case, and Eq. (\ref{Eq:pmcratio}) takes a large value. 
Hence, the nonreciprocity from the parity mixing is dominant.
However, in the case of $\EF < 0$, the nonreciprocal paraconductivity is from the cubic term mechanism because the parity mixing does not create nonreciprocity in that regime \cite{WakatsukiNagaosa}.

\subsection{Surface state of topological insulator} \label{sec:TI_sc}
\subsubsection{Parabolic dispersion}
We consider the nonreciprocal paraconductivity in the superconducting surface state of a topological insulator.
The simplest Hamiltonian for the surface state is
\begin{equation}
H = v_{\rm F} \left(k_x \sigma_y - k_y \sigma_x\right) - \bB \cdot 
\bm{\sigma},
\end{equation}
with $v_{\rm F}$ ($>0$) being the Fermi velocity and the magnetic field is applied along the in-plane direction.
However, the in-plane magnetic field simply shifts the momentum and does not affect the transport.
Therefore, additional terms are necessary for the nonreciprocal current.

In this subsection, we include the term proportional to $\bk^2$, 
whose Hamiltonian form
is equivalent to Eq. (\ref{Eq:rashba}), with $\alpha$ replaced with the Fermi velocity $v_{\rm F}$.
However, the first (parabolic) term here is much smaller than the second ($\bm k$-linear) term.
Therefore, we will consider the asymptotic form for 
large $m$.
Furthermore, we take into account only the inner branch of the Fermi surfaces.
The eigenenergies are
\begin{equation}
\xi_{\bk} = \frac{\bk^2}{2m} - \EF \pm \sqrt{\left(v_{\rm F} k_x - B_y\right)^2 
+ \left(v_{\rm F} k_y\right)^2}, \label{Eq:rashbaband}
\end{equation}
with the Fermi energy $\EF$ included.
We first assume that the Fermi energy is on the conduction band (plus sign in the equation and 
$\EF > 0$).
We calculate the GL free energy corresponding to Eq. (\ref{Eq:rashbaband}).
The detailed derivation is shown in Appendix A, and the result is 
\begin{widetext}
\begin{equation}
F = \int \frac{d^2 \bq}{\left(2\pi\right)^2} \left[
\frac{1}{g} - \nu_1 S_1 \left(T\right) + \nu_1 S_3\left(T\right) \left( 
\frac{\left(\kFa+m v_{\rm F}\right)^2}{8m^2}\bq^2 + 
\frac{3\left(5\kFa+3m v_{\rm F}\right)}{32m\kFa\left(\kFa+m v_{\rm F}\right)}B_y q_x 
\bq^2 \right) \right] \left|\Psi_{\bq}\right|^2. 
\label{Eq:free2}
\end{equation}
\end{widetext}
We can also show that the free energy in the case where $E_{\rm F}$ is located on the valence band ($E_{\rm F}<0$) is obtained by 
replacing $v_{\rm F}$ with $-v_{\rm F}$, $B_y$ with $-B_y$, $\nu_1$ with $\nu_2$, and $\kFa$ with $\kFb$. It is noted that the cubic term with respect to the wavenumber vanishes for 
$m \rightarrow \infty$, which is consistent with the fact that the in-plane magnetic 
field simply shifts the momentum of the Cooper pairs without the parabolic term.

The paraconductivity up to $O\left(E_x^2 B_y\right)$ can be obtained by applying 
Eq. (\ref{Eq:paracon}) as
\begin{equation}
j_x = \frac{e^2}{32\epsilon}E_x - \frac{3\pi e^3 m \left(5\sqrt{m\EFR}-2m v_{\rm F}\right)
B_y}{4096\left(m\EFR\right)^{3/2}\left(\sqrt{m\EFR}-m v_{\rm F}\right) T_0 \epsilon^2}E_x^2. 
\label{Eq:jx}
\end{equation}
The temperature dependence of the conductivity is the same as those of the transition metal dichalcogenides and the Rashba superconductors \cite{Wakatsuki, WakatsukiNagaosa}.
We can also show that the case for $\EF < 0$ has the same form,
and
the sign of the nonreciprocal current compared to the $\EF>0$ case is reversed.

The corresponding $\gamma$-value is
\begin{equation}
W \gamma_\mathrm{S} = -\frac{3\pi m 
\left(5\sqrt{m\EFR} - 2m v_{\rm F}\right)}{4e\left(m\EFR\right)^{3/2}
\left(\sqrt{m\EFR}-m v_{\rm F}\right)T_0},
\label{eq:TI_1}
\end{equation}
and for $m \rightarrow \infty$
\begin{equation}
W \gamma_\mathrm{S} \rightarrow - \frac{9 \pi}{4 e m v_{\rm F} \EF T_0}
,
\end{equation}
which is a leading contribution in the limit of a small parabolic term.

Now, we compare the $\gamma$-value with that of the normal state.
The normal state conductivity is calculated by the Boltzmann equation with the relaxation time approximation \cite{IdeueHamamoto}:
\begin{align}
j_x &= - \tau e^2 E_x \int \frac{d^2\bk}{\left(2\pi\right)^2} v_{\bk} \partial_{k_x} f_{\bk}
 - \tau^2 e^3 E_x^2 \int \frac{d^2\bk}{\left(2\pi\right)^2} v_{\bk} \partial_{k_x}^2 f_{\bk} 
, \label{eq:Boltzmann}
\notag \\
&\hspace{-5mm}
= \tau e^2 E_x \int \frac{d^2\bk}{\left(2\pi\right)^2} v_{\bk}^2 \delta\left(\xi_{\bk}\right) 
  -\tau^2 e^3 E_x^2 \int \frac{d^2\bk}{\left(2\pi\right)^2} \partial_{k_x} v_{\bk} v_{\bk} 
\delta\left(\xi_{\bk}\right),
\end{align}
where $\xi_{\bk}$ is the eigenenergy,
$v_{\bk} = \frac{\partial \xi_{\bk}}{\partial k_x}$ is the group velocity, and $\tau$ is the relaxation time.
The Fermi distribution function is also defined by $f_{\bm k} = 1/ (e^{\beta\xi_{\bm k}}+1)$.
With use of this, we obtain the electronic current
\begin{equation}
j_x = \frac{e^2 \tau \left(\EFR - \sqrt{m v_{\rm F}^2\EFR}\right)}{4\pi}E_x 
- \mathrm{sgn}\left(\EF\right)\frac{3 e^3 \tau^2 B_y}{16\pi \sqrt{m\EFR}}E_x^2,
\end{equation}
The $\gamma$-value is
\begin{equation}
W \gamma_\mathrm{N} = -\mathrm{sgn}\left(\EF\right)\frac{3 \pi}{e\sqrt{m\EFR}
\left(\EFR-\sqrt{m v_{\rm F}^2 \EFR}\right)^2},
\end{equation}
and for $m \rightarrow \infty$
\begin{equation}
W \gamma_\mathrm{N} \rightarrow - \mathrm{sgn}\left(\EF\right)
\frac{3 \pi}{e m v_{\rm F} \EF^2}.
\label{eq:TI_parab_gamma_N}
\end{equation}
Therefore, the ratio for $m \rightarrow \infty$ is
\begin{equation}
\frac{\gamma_\mathrm{S}}{\gamma_\mathrm{N}} \rightarrow 
\frac{3 \left|\EF\right|}{4T_0},
\label{eq:TI_parab_ratio}
\end{equation}
which does not depend on $m$ and $v_{\rm F}$.
The enhancement of the nonreciprocity by the factor of $\EF / T_0$ is expected in the 
superconducting fluctuation regime, although the power of enhancement is different from the 
TMD \cite{Wakatsuki}, in which $\gamma_\mathrm{S} / 
\gamma_\mathrm{N} \sim \left(\EF / T_0\right)^3$.
It should be noted that the direction of the nonreciprocal current is reversed between the 
conduction and valence bands in both of the paraconductivity and normal conductivity.

\subsubsection{Hexagonal warping}
The surface band of topological insulators such as Bi$_2$Te$_3$ or Bi$_2$Se$_3$ is 
hexagonally distorted because of the crystal symmetry \cite{Fu}.
We consider the effect of the hexagonal warping on the nonreciprocal paraconductivity.
The Hamiltonian is
\begin{align}
H = v_{\rm F} \left(k_x \sigma_y - k_y \sigma_x\right) + \frac{\sqrt{\lambda}}{2} \left(k_+^3 
+ k_-^3\right) \sigma_z - B_y \sigma_y, \label{Eq:hexagonal}
\end{align}
with $k_\pm = k_x \pm i k_y$ and $\lambda$ describes the strength of the hexagonal warping.
The corresponding GL free energy can be obtained in a similar manner to the previous subsection.
The free energy up to $O(\bq^3 B_y \lambda)$ is
\begin{widetext}
\begin{align}
F = \int \frac{d^2 \bq}{\left(2\pi\right)^2} \left[
\frac{1}{g} - \nu S_1 \left(T\right) + \nu S_3\left(T\right) \left( \left(\frac{v_{\rm F}^2}{8} 
+ \frac{5 \lambda \kF^4}{16} \right) \bq^2 + \frac{37 \kF^2 \lambda}{8v_{\rm F}} 
B_y q_x \bq^2 \right)
\right] \left|\Psi_{\bq}\right|^2,
\label{eq:free_energy_TI_hexa_cubic}
\end{align}
\end{widetext}
with $v_{\rm F} \kF = |\EF|$ and the density of states $\nu$ at the Fermi level.
The paraconductivity is obtained as
\begin{align}
j_x = \frac{e^2}{32\epsilon}E_x - 
\frac{37 \pi e^3 \EF^2 \lambda B_y}{1024 v_{\rm F}^5 T_0 \epsilon^2}E_x^2.
\end{align}
Consequently, the $\gamma$-value is
\begin{align}
W \gamma_\mathrm{S} = -\frac{37 \pi \EF^2 \lambda}{e v_{\rm F}^5 T_0}.
\label{eq:TI_2}
\end{align}

We also calculate the normal state current based on Eq.~\eqref{eq:Boltzmann}.
We express $\bk$ in the polar coordinate $(k, \theta)$, and solve $\xi_{\bk} = 0$ up to 
$O(B_y \lambda)$ with fixed $\theta$.
Then, the integrals over $k$ and $\theta$ can be carried out.
The result is
\begin{equation}
j_x = \frac{\pi \tau e^2 |\EF| \left(1 + \EF^4 \lambda / v_{\rm F}^6\right)}{4\pi^2} E_x 
- \frac{9 \tau^2 e^3 |\EF|^3 \lambda B_y}{2\pi v_{\rm F}^5}E_x^2.
\end{equation}
The $\gamma$-value is
\begin{equation}
W \gamma_\mathrm{N} = -\frac{72 \pi |\EF| \lambda}{e v_{\rm F}^5}.
\label{eq:}
\end{equation}
Therefore, the ratio of the $\gamma$-value of the paraconductivity and normal 
conductivity is
\begin{equation}
\frac{\gamma_\mathrm{S}}{\gamma_\mathrm{N}} = \frac{37 |\EF|}{72 T_0},
\end{equation}
which again does not depend on $v_{\rm F}$ and $\lambda$.
The order of magnitude of the enhancement factor is the same as that for the parabolic dispersion case.
We emphasize that the direction of the nonreciprocal current does not change between the 
conduction and valence bands in both of the paraconductivity and normal conductivity, which is 
clearly different from the nonreciprocal current originates from the parabolic term in the 
previous subsection.

\subsection{Transition metal dichalcogenides} \label{sec:TMD}
The nonreciprocal paraconductivity in two-dimensional TMD has 
been investigated both theoretically and experimentally \cite{Wakatsuki}.
Here we summarize these results.
The normal state Hamiltonian around the $K$ and $-K$ valleys is
\begin{equation}
H_{\bk\sigma\tau} = \frac{\bk^2}{2m} + \tau_z \lambda k_x 
\left(k_x^2 - 3k_y^3\right)-\DZ \sigma_z - \DSO \sigma_z \tau_z, \label{Eq:tmd}
\end{equation}
with $\bk$, $m$, $\lambda$, $\DZ=B_z$, and $\DSO$ being the electron wavenumber, mass, 
amplitude for the trigonal warping, Zeeman splitting, and spin-orbit splitting, respectively.
Here $x$-axis is taken along a `zigzag' chain in the honeycomb lattice, and $y$-axis is along `armchair' direction.
The out-of-plane magnetic field is necessary for the nonreciprocal current.
Here, $\sigma_z$ and $\tau_z$ represents the Pauli matrices for the spin and valley degrees of 
freedom, respectively.
The GL free energy for the superconducting state can be derived based on Eq. (\ref{Eq:tmd}).
The GL free energy up to $O\left(\DZ \lambda {\bm q}^3\right)$ is
\begin{equation}
F = \int \frac{d^2 \bm q}{(2\pi)^2} 
\, \Psi_{\bm q}^* \left[ a + \frac{\bm q^2}{4m} + \Lambda B_z \left(q_x^3 
- 3q_xq_y^2\right) \right] \Psi_{\bm q}
, \label{Eq:Ftmd}
\end{equation}
with 
$\Lambda = \frac{93 \zeta\left(5\right)}{14\zeta\left(3\right)} 
\frac{\DSO\lambda}{\left(\pi T_0\right)^2}$.
The effect of nonmagnetic impurities can also be considered, whose derivation is summarized in Appendix B.
By applying Eq. (\ref{Eq:paracon}) for Eq. (\ref{Eq:Ftmd}), the current up to the second order of electric field is obtained as
\begin{equation}
\bj = \frac{e^2}{16\epsilon} \bE - 
\frac{\pi e^3 m \Lambda B}{64T_0 \epsilon^2}\bF\left(\bE\right)
,\label{eq:TMD_current}
\end{equation}
with $\bF\left(\bE\right) 
= \left(E_x^2-E_y^2, -2E_xE_y\right)$, which is 
consistent with the crystal symmetry of the transition metal dichalcogenides.
The $\gamma$-value is
\begin{equation}
W \gamma_\mathrm{S} = \frac{4\pi m\Lambda}{e T_0}.
\label{eq:gamma_TMD}
\end{equation}
On the other hand, the $\gamma$-value for the normal state can be obtained by the Boltzmann 
equation \cite{IdeueHamamoto}, whose typical value is calculated as $W\gm_{\rm N}\sim m\Delta_{\rm SO}\lambda/(eE_{\rm F}^3)$ \cite{Wakatsuki}.
The ratio between the superconducting fluctuation 
and normal regimes is found to be $\gamma_\mathrm{S} / \gamma_\mathrm{N} 
\sim \left(\EF / T_0\right)^3$, which is quite large.
While the focus of this paper is on the region at small magnetic fields, we can also consider the high magnetic field region by including the Landau level, which is discussed in Appendix C.

\section{Vortex in noncentrosymmetric superconductors
} \label{Sec:Vortex}

When the superconducting gap function is sufficiently developed below $T_0$, the phase of the order parameter forms the vortices whose dynamics governs electronic transport properties.
In this section, we begin with the phenomenological discussion for the nonreciprocity entering through the renormalization of the superfluid density, which captures the essence of the nonreciprocal vortex dynamics.
In Sec.~\ref{Subsec:Ratchet}, we also consider the the ratchet potential effect for the vortices, which is necessarily present for noncentrosymmetric superconductors with trigonal symmetry and disorder effects.

\subsection{Renormalization of superfluid density} \label{sec:renormalization}

\subsubsection{Modified KT transition point for a system with in-plane magnetic field} \label{sec:KT_transition}
Let us first consider the system with the in-plane magnetic field.
In this case, the vortices and antivortices are generated thermally above the KT transition temperature.
We take a Rashba ($E_{\rm F}<0$) or TI-based system with in-plane magnetic field, whose free energy density is in general given by
\begin{align}
f &= \tfrac 1 2 \rho_s \bm v_s^2 + \Lambda' B_y \rho_s v_{sx}(v_{sx}^2 + v_{sy}^2)
,
\end{align}
which is phenomenologically introduced by utilizing the replacement $\bm q \rightarrow m^*\bm v_s$ in the GL theory.
Here $\rho_s =m^* n_s$ and $\bm v_s$ are superfluid mass density and velocity, respectively.


We use a mean-field approach for analysis, in which the terms higher than second-order are approximated into quadratic one.
Under the externally induced current, the superfluid acquires the uniform velocity component 
$\bm v_{\rm unif}$.
Hence we can replace one of $\bm v_s$ by $\bm v_{\rm unif}$ in the third-order term
 and obtain the free energy density
\begin{align}
f &\simeq \tfrac 1 2 \sum_{\mu\nu} \tilde \rho_{s,\mu\nu} v_{s\mu} v_{s\nu}
,\\
\tilde \rho_s &=  \rho_s
\begin{pmatrix}
1+6\Lambda' B_y v_{{\rm unif},x} & 2\Lambda' B_y v_{{\rm unif},y} \\
2\Lambda' B_y v_{{\rm unif},y}  & 1-2\Lambda' B_y v_{{\rm unif},x}
\end{pmatrix}
.
\end{align}
Thus the uniform current renormalizes the superfluid density 
due to the presence of the $\bm v$-cubic term.
If we choose $\bm v_{\rm unif} = v_{s0} \hat{\bm x}$, the free energy has the following form
\begin{align}
f&\simeq \tfrac 1 2 \rho_s (1+4\Lambda' B_y v_{s0}) (v_{sx}^2 + v_{sy}^2)
\nonumber \\
&\ \ \ 
+\rho_s  \Lambda' B_y v_{s0} (v_{sx}^2 - v_{sy}^2)
. \label{eq:free_2x}
\end{align}
The first term here renormalizes the superfluid 
density isotropically:
\begin{align}
\tilde n_s = n_s (1+4\Lambda' B_y v_{s0})
. \label{eq:ns_renorm}
\end{align}
The second term in Eq.~\eqref{eq:free_2x} is the anisotropic renormalization, which will be discussed in detail in the next subsections.
In the following, we follow the procedure given in Ref.~\onlinecite{Halperin79}.

The isotropic renormalization of the superfluid density has an influence on the force between 
the two vortices.
This force can be interpreted as the Magnus force acting on one vortex which is located in a 
nearly uniform current made by the other vortex.
At KT transition point, this force is balanced by 
an entropic force proportional to temperature,
which leads to 
the force balance relation for two vortices separated by the sufficiently large distance $r$ \cite{Nelson77,Halperin79}:
\begin{align}
e^* \tilde n_s \frac{1}{m^*r} \times \Phi_0 = \frac{4\tilde T_{\rm KT}}{r}
, \label{eq:KT_balance}
\end{align}
where $\Phi_0 = 2\pi/|e^*|$ is the magnetic flux quantum and $e^*=2e$.
Thus, the important consequence of the isotropically renormalized superfluid density is the modification 
of the KT transition temperature which has different values depending on the direction of the 
uniform current.

Through the modification of $T_{\rm KT}$, the correlation length $\xi_+$, which is exponentially diverging near $T_{\rm KT}$, 
and the density of unpaired vortices $n_{\rm v}$ are also affected by $\bm v_{{\rm unif}}$.
These physical quantities are given near $T_{\rm KT}$ by \cite{Kosterlitz74,Halperin79}
\begin{align}
\xi_+ &= 
b^{-1/2}
\xi_c \exp \left(
\sqrt{b \frac{T_0 - \tilde T_{\rm KT}}{T-\tilde T_{\rm KT}}}
\right)
,
\\
n_{\rm v} &= (2\pi C_1 \xi_+^{2})^{-1}
,
\end{align}
where $
b,C_1$ are order of unity constants, and $\xi_c$ is the GL coherence length evaluated 
at $T=T_{\rm KT}$.
The mean-field transition temperature is written as $T_0$.
The value of $n_{\rm v}$ goes to zero at $\tilde T_{\rm KT}$ with only bounded vortices left.
Since the correlation length is dependent of the KT transition temperature, this indicates that 
the number of vortices is different depending on the direction of the uniform current.

Now we consider the electric field caused by the vortex dynamics \cite{Halperin79}.
With a uniform current, the thermally generated vortices feel a Magnus force and move with the 
velocity $\pm \bm v_{\rm L}$ perpendicular to $\bm v_{\rm unif}$.
To convert a hydrodynamic effect to electric one,
we use the Josephson relation $\varDelta V = (1/e^*) {d}\varDelta \theta/{d t}$ 
where $\varDelta V$ and $\varDelta \theta$ are voltage drop and phase difference between the 
two ends of the sample, respectively.
The vortices of the number $n_{\rm v} L_x v_{\rm L}$ with the sample width $L_x$ along $x$-direction cross the sample edges per 
unit time, leading to the phase slip $d \varDelta \theta/d t = 2\pi n_{\rm v} L_x v_{\rm L}$.
The electric field $E_x = \varDelta V/L_x$ is then given by
\begin{align}
E_x = \frac{(2\pi)^2 n_{\rm v} }{|e^*|^2\eta} j_{{\rm unif},x}
, \label{eq:Josephson_relation}
\end{align}
where we have used the force balance relation $j_{{\rm unif},x}\Phi_0 = \eta v_{\rm L}$ between 
Magnus force and friction force.
The uniform electric current has been defined by 
$\bm j_{\rm unif}= -\partial f/\partial \bm A = \frac{e^*}{m^*} \partial f/\partial \bm v_{\rm unif} = e^* n_s \bm v_{\rm unif} + O(\bm v_{\rm unif}^2)$, whose higher-order terms do not modify the conclusion for the $\gm$-value.
The friction coefficient is given by $\eta = \eta_0 (1+d_0\Lambda' B_y v_{s0})$ with $d_0$ being a constant and $\eta_0 \simeq \pi\sg_n (e^*)^2 \xi^2 $.
This modification enters due to the anisotropic renormalization of superfluid density.
The derivation is given in Secs.~III\,A\,2 and III\,B in detail, and here we just employ the final results.
Substituting these expressions, we obtain the resistivity
\begin{align}
\rho = \frac{2(1-d_0\Lambda' B_y v_{s0})}{C_1\sg_n}
\left(
\frac{\xi_c}{\xi_+}
\right)^{2}
.
\end{align}
Thus, there are two kinds of the sources for nonreciprocal response: one is from the modified KT transition temperature and the other from the modified friction coefficient.
By extrapolating the above expression to $T\rightarrow T_0$, the exponential temperature dependence in the correlation length is not effective.
In this case, with the transport coefficients defined in $E_x = \rho_1 j_{{\rm unif},x} + \rho_2 j_{{\rm unif},x}^2$, we obtain the explicit expressions
\begin{align}
\rho_1 &= \frac{2b}{C_1\sg_n}
, \\
\rho_2(T) &= - \frac{d_0\Lambda' B_y}{e^* n_s(T)} \rho_1
.
\end{align}
We then find that the $\gm$-value, which is given by $\gm_{\rm S}^{\rm v} = - \rho_2/\rho_1B_y W$, is proportional to $(T_0-T)^{-1}$, where we have used the temperature dependence of the superfluid density as $n_s(T) \propto T_0 - T$.

We now switch our discussion to the lower temperatures near the KT transition.
We write the KT transition temperature as $\tilde T_{\rm KT} = T_{\rm KT} + \delta T_{\rm KT}$ with 
the unrenormalized transition temperature $T_{\rm KT}$, and expand the expression with 
respect to $\delta T_{\rm KT}$.
We assume $T_0 - T_{\rm KT} \gg \delta T_{\rm KT}$ and then the transport coefficients are given 
for the temperature range $T - T_{\rm KT} \gg \delta T_{\rm KT}$ by
\begin{align}
\rho_1(T) &= \frac{2b}{C_1\sg_n} \exp \left(
-2
\sqrt{b \frac{T_0 - T_{\rm KT}}{T-T_{\rm KT}}}
\right)
, \\
\rho_2(T) &= - \frac{2\pi\Lambda' B_y 
\sqrt{b(T_0-T_{\rm KT})}}
{m^*e^*(T-T_{\rm KT})^{3/2}}\,\rho_1(T)
,
\end{align}
where we have kept the leading order contribution remaining for $T\rightarrow T_{\rm KT}$.
While both $\rho_1$ and $\rho_2$ exponentially goes to zero toward $T_{\rm KT}$,
the $\gm$-value is proportional 
to $(T-T_{\rm KT})^{-3/2}$.
Thus a large nonreciprocal signal is expected near the KT transition point.

Below $T_{\rm KT}$, the linear response vanishes due to bounded vortices, and instead the third-order term characterizes the current-voltage relation.
The nonreciprocal response should then be reflected in the fourth-order term.
Hence the $I$-$V$ relation has the form $V = a_3 I^3 + a_4 I^4$ with $a_4 = a_4' B$ near $T_{\rm KT}$.
The higher-order terms become more relevant at lower temperatures.

\subsubsection{Extended Bardeen-Stephen approach for a system with out-plane magnetic field}
While the vortices in a system with in-plane magnetic field are created by the thermal fluctuation, the out-plane magnetic field creates the vortices having the same vorticity, which is qualitatively different situation from the in-plane case.
We here extend the Bardeen-Stephen theory \cite{Bardeen65,Tinkham_book} for
 flux-flow conductivity to non-linear response regime.
We take the MoS$_2$-based system where the magnetic field is applied along out-plane direction.
The nonreciprocity for this system is considered as an effective renormalization of the superfluid density.
To see this, we begin with the free energy density for a superfluid
\begin{align}
f &= \tfrac 1 2 \rho_s \bm v_s^2 + \Lambda' B_z \rho_s v_{sx}(v_{sx}^2 - 3v_{sy}^2)
.
\end{align}
By comparing this expression with Eq.~\eqref{Eq:Ftmd}, we find the relations $m^*=2m$ and $\Lambda' \sim m^2 \Lambda$.
We choose the uniform current flowing along $x$-direction: 
$\bm v_{\rm unif} = v_{s0} \hat{\bm x}$.
The free energy then becomes
\begin{align}
f &= \tfrac 1 2 (\tilde \rho_{s,xx} v_{sx}^2 + \tilde \rho_{s,yy} v_{sy}^2)
,\label{eq:f_dens}
\\
\tilde \rho_{s,xx} &= \rho_s (1+6\Lambda' B_z v_{s0})
,\\
\tilde \rho_{s,yy} &= \rho_s (1-6\Lambda' B_z v_{s0})
,
\end{align}
within the `mean-field' approximation explained above.
Here only the anisotropic renormalization occurs.
Using the velocity potential, the superfluid velocity can be written as 
$\bm v_s = \frac{1}{m^*}\bm \nabla \theta$.
Let us first consider the isolated superconducting vortex induced by the out-plane external 
magnetic field.
The vortex velocity for isotropic system is given by $\theta = \phi = \tan^{-1}(y/x)$ with the 
polar coordinate $\bm r=(r,\phi)$, which satisfies the equation 
$\bm \nabla ^{2}\theta = 0$ obtained from a variational principle.
We can make Eq.~\eqref{eq:f_dens} isotropic by the scaling 
$x' = x\sqrt{\rho_s/\tilde \rho_{s,xx}}$ and 
$y' = y\sqrt{\rho_s/\tilde \rho_{s,yy}}$.
In this case, the velocity potential is given by $\theta = \tan^{-1}(y'/x')$.
Keeping the first-order contribution with respect to $\Lambda'$, we get
\begin{align}
\theta(\phi) = \phi + 3\Lambda' B_z v_{s0} \sin 2\phi
. \label{eq:potential_mod}
\end{align}
In order to calculate the spatial distribution of electric field, we employ
the London equation outside the normal core accounting for zero resistivity: 
$\bm E = \varLambda \partial_t \bm j_s$ where $\varLambda^{-1} = n_s (e^*)^2 /m^*$ \cite{Bardeen65,Tinkham_book}.
For a moving vortex with the velocity $\bm v_{\rm L}$, the spatial coordinate is replaced as 
$\bm r\rightarrow \bm r -\bm v_{\rm L} t$ and we can replace the time-derivative as 
$\partial_t \rightarrow -\bm v_{\rm L}\cdot\bm \nabla$.
With a uniform current along $x$-direction, the ``Lorentz'' force acting on the vortex is along 
$y$-direction, so we choose $\bm v_{\rm L} = v_{\rm L} \hat{\bm y}$.
(Since the magnetic penetration depth is very long for atomically thin two-dimensional 
superconductors, the force should originate from fluid-mechanical Magnus force \cite{Nozieres66}, although the main 
conclusion in the following is not altered.)
Substituting $\bm j_s= e^* n_s\bm v_s$ into the London equation, we can calculate the electric 
field in the superconducting region, where $r$ is larger than the coherence length $\xi$, as
\begin{align}
\bm E(r>\xi) &= \frac{v_{\rm L}}{e^*r^2} \cos \phi (1+6\Lambda' B_z v_{s0} \cos 2\phi) \bm e_r
\nonumber \\
&\ \ + \frac{v_{\rm L}}{e^*r^2} \sin \phi [1+6\Lambda' B_z v_{s0}(2+3 \cos 2\phi)] \bm e_\phi
.
\end{align}
For the inside of the normal core ($r<\xi$), the electric field $\bm E = - \bm \nabla \varphi$ 
is determined from the Poisson equation $\bm \nabla^2 \varphi=0$ due to a charge neutrality.
We assume that the boundary between superconducting and normal regions is circular at $r=\xi$ 
and is connected discontinuously.
With this geometry only the $\bm e_\phi$ component of $\bm E$ matters for the boundary condition.
We expand the scalar potential as $\varphi = \sum_{m} C_m (r\epn^{\imu \phi})^m$ and 
the boundary condition gives 
\begin{align}
\varphi(r<\xi) &= -\frac{v_{\rm L}}{e^*\xi^2}(1+3\Lambda' B_z v_{s0}) r \cos \phi 
\nonumber \\
&\ \ \ 
- \frac{3v_{\rm L}\Lambda' B_z v_{s0}}{e^*\xi^4} r^3 \cos 3\phi
.
\end{align}
The energy dissipation rate is then calculated as
\begin{align}
W(r<\xi)=
\sg_n \bm E^2
= \frac{\sg_n }{(e^*)^2 \xi^4} (1+6\Lambda' B_z v_{s0}) v_{\rm L}^2
, \label{eq:dissipation_BS}
\end{align}
where $\sg_n$ is a normal conductivity.
We note that this expression does not depend on the spatial coordinate $\bm r$ in the leading order.
Equation~\eqref{eq:dissipation_BS} is identified as the energy dissipation rate per unit volume in 
the form $\eta v_{\rm L}^2/ \pi \xi^2$ originating from the friction force for the vortex flow.
We thus arrive at an important conclusion that the effect of the cubic term in free energy is reflected 
in the change of the friction coefficient $\eta$.

The ``Lorentz'' force is balanced by the friction force as
$(\bm j_{\rm unif} \times \bm \Phi_0)_y = \eta v_{\rm L}$
where $\eta = \eta_0 (1+6\Lambda' B_z v_{s0})$ with $\eta_0 = \pi \sg_n/(e^*)^2\xi^2$ and 
we have defined $\bm \Phi_0 = \Phi_0 \hat{\bm z}$.
We use
the expressions for the uniform current $\bm j_{\rm unif} = e^*n_s \bm v_{\rm unif}$ and 
the "Faraday's law'' $\bm E_0 = \bm B\times \bm v_{\rm L}$ for the uniform electric field generated from 
the 
flux flow.
(Since the magnetic field $\bm B$ is not time-dependent in the system with infinite magnetic penetration depth, this ``Faraday's law'' does not derive from an electromagnetic origin but from the Josephson relation as in the previous subsection.)
We then obtain $E_{0x} = \rho_1 j_{{\rm unif},x} 
+ \rho_2 j_{{\rm unif},x}^2$ where
\begin{align}
\rho_1 &= \frac{B_z\Phi_0}{\eta_0}
, \\
\rho_2 &= - \frac{6\Lambda' B_z^2\Phi_0}{e^* n_s \eta_0}
.
\end{align}
The $\gm$-value from the vortex dynamics is given by
\begin{align}
\gm_{\rm S}^{\rm v} &= - \frac{\rho_2}{\rho_1B_z W} = \frac{6\Lambda'}{e^* n_s W}
,
\end{align}
which is not dependent on the normal conductivity.
Rewriting the coefficient $\Lambda'$ in terms of the quantities in Sec.~\ref{sec:TMD},
we can estimate
the ratio between $\gm_{\rm S}^{\rm v}$ from vortex motion and $\gm_{\rm S}^{\rm f}$ from 
superconducting fluctuation as
\begin{align}
\frac{\gm_{\rm S}^{\rm v}}{\gm_{\rm S}^{\rm f}} \sim \frac{m T_0}{n_s}
.
\end{align}
Using the relations $2n_s/n_e \simeq (T_0 - T)/T_0$ and the normal electron density 
$n_e = k_{\rm F}^2/2\pi$, we finally obtain
\begin{align}
\frac{\gm_{\rm S}^{\rm v}}{\gm_{\rm S}^{\rm f}} \sim \frac{T_0^2}{E_{\rm F} (T_0-T)}
\xrightarrow{\ T\rightarrow0\ }
 \frac{T_0}{E_{\rm F}}
.
\end{align}
It is characteristic that the $\gm$-value from vortex dynamics is enhanced near the transition 
temperature and has a smaller value with the factor $T_0/E_{\rm F}$ compared to $\gm_{\rm S}^{\rm f}$ 
at low temperatures.
The same conclusion will be obtained from the TDGL approach as described in Sec.~\ref{sec:TDGL} 
in detail, and hence the above approach can be justified.

\subsection{TDGL approach} \label{sec:TDGL}
\subsubsection{Formulation}
The vortex dynamics induced by the out-plane magnetic field can be described by the 
TDGL theory, which takes account of a purely dissipative dynamics.
This approach is successfully applied to the flux-flow conductivities \cite{Schmid66,Kopnin_book}.
While this theory is less intuitive than the above extended Bardeen-Stephen theory, the 
TDGL approach gives a foundation for its interpretation.
Here we formulate the theory by including the cubic term in $\bm q$ responsible for a 
nonreciprocal vortex dynamics.
We begin with the TDGL equation and free energy
\begin{align}
&\Gamma \left( \partial_t + 2\imu e\varphi \right) \Delta = 
- \frac{\delta F}{\delta \Delta^*}
, \label{eq:TDGL_vtx} \\
&F_s =\int \diff \bm r\, \Delta^*\left[
\al + \frac{\beta}{2} |\Delta|^2 + \gm \bm P^2 + K (P_x^3-3\overline{P_xP_y^2})
\right] \Delta 
\nonumber \\
&\hspace{7mm}
+ \frac{1}{2\mu_0} \int\diff \bm r  \bm B^2
, \label{eq:TDGL_Fs}
\end{align}
where $\bm P = - \imu \bm \nabla - 2e\bm A$.
The gap parameter is related to the wave function by $\Psi = \sqrt{\frac{7\zeta(3)n_e}{2\pi^2 T_0^2}}\, \Delta$ with the electron number $n_e$.
There are also the relations $K \propto \Lambda B_z$ and $\al \propto (T_0 - T)$.
Note that the coefficient $\gm$ in Eq.~\eqref{eq:TDGL_Fs} is different from the $\gm$-value for nonreciprocal tansport which is denote as $\gm_{\rm S}$.
The total GL free energy is given by $F = F_s + F_n$ with $F_n$ being a normal part.
We have defined the symmetrization by
\begin{align}
\overline{ABC} &= \frac{1}{3!}
(ABC+BCA+CAB+ACB+BAC+CBA)
,
\end{align}
to make the free energy real.
The supercurrent density is given by $\delta F / \delta \bm A = 0$ as 
\begin{align}
&\hspace{-5mm} j_{sx} 
= 2e\gm
(\Delta^* P_x \Delta +  P_x^\dg \Delta^* \Delta)
+  2eK \left(
\Delta^*P_x^2 \Delta 
+P_x^\dg\Delta^*P_x \Delta 
\right.
\nonumber \\
&\ \ \ 
\left.
+ P_x^{\dg2}\Delta^* \Delta 
- \Delta^*P_y^2 \Delta 
-P_y^\dg\Delta^*P_y \Delta 
- P_y^{\dg2}\Delta^* \Delta 
\right)
, \\
&\hspace{-5mm}j_{sy} 
= 2e\gm
(\Delta^* P_y \Delta +  P_y^\dg \Delta^* \Delta)
- 2eK \left(
\Delta^*P_xP_y \Delta 
+\Delta^*P_yP_x \Delta 
\right.
\nonumber \\
&\ \ \ 
\left.
+P_x^\dg\Delta^*P_y \Delta 
+P_y^\dg\Delta^*P_x \Delta 
+ P_x^{\dg} P_y^{\dg}\Delta^* \Delta 
+ P_y^{\dg} P_x^{\dg}\Delta^* \Delta 
\right)
.
\end{align}
The total current is given by $\bm j = \bm j_n + \bm j_s$ with the normal current 
$\bm j_n = \sg_n \bm E$.
We can also show the relations
\begin{align}
\frac{\delta F}{\delta \Delta^*} 
&= \left[
\al + \beta |\Delta|^2 + \gm \bm P^2 + K (P_x^3-3\overline{P_xP_y^2})
\right] \Delta
,
\end{align}
and
\begin{align}
\imu \left(
\Delta \frac{\delta F}{\delta \Delta}  - \Delta ^*\frac{\delta F}{\delta \Delta^*} 
\right) = - \frac{1}{2e} \bm \nabla \cdot \bm j_s
.
\end{align}
The above expressions can be used for arbitrary strength of the magnetic field.

In the following, we concentrate on the two-dimensional superconductor where the 
magnetic penetration depth is typically longer than the sample size \cite{Pearl64}.
In this case, the magnetic effects can be neglected and only the electric and fluid mechanical 
effects are considered in the GL analysis.
Such condition can be set by choosing $\bm A=0$.

With a slightly shifted center as $\bm r\rightarrow \bm r+\bm d$, we have the relation
\begin{align}
\delta F &= \int \diff \bm r \left[
 \frac{\delta F}{\delta \Delta}   (\bm d \cdot \bm \nabla) \Delta
+{\rm c.c.}
\right]
,
\end{align}
which defines a friction force acting on superconductor.
This force is balanced by an external force to result in a stationary motion of the vortex.
The force balance equation for an isolated  single vortex is given by
\begin{align}
\bm j_{\rm unif} \times {\bm \Phi}_0
&= - \Gamma \int \diff \bm r
\left[
  \left(\partial_t - 2\imu e\varphi \right) \Delta^* \bm \nabla \Delta 
+ {\rm c.c.}
\right]
.
\end{align}
The transport current from an external source is written as $\bm j_{\rm unif}$.
The term on the left-hand side is the external force acting on the fluxoide, which is in 
general composed of the sum of Lorentz and Magnus forces \cite{Kato16}.
In the current situation, only the Magnus force contributes \cite{Nozieres66} since the 
London magnetic penetration depth is taken as infinity.

We also need the equation for the scalar potential, for which the equation of continuity 
$\bm \nabla\cdot \bm j= 0$ 
is used.
The explicit form is given by
\begin{align}
\left(
\frac{\sg_n}{2e} \bm \nabla^2 - 4e\Gamma |\Delta|^2
\right)\varphi = \imu \Gamma (\Delta \partial_t \Delta^* - \Delta^* \partial_t \Delta)
,
\end{align}
where we have used $\bm E = - \bm \nabla \varphi$.

For a moving vortex, the spatial coordinate can be written in the laboratory frame by the 
replacement $\bm r\rightarrow \bm r - \bm v_{\rm L} t$ with a boost velocity $\bm v_{\rm L}$ for the vortex.
Correspondingly, we replace the time-derivative as 
$\partial_t \rightarrow - \bm v_{\rm L}\cdot \bm \nabla$ and rewrite the equations with 
dimensionless quantities as
\begin{align}
&v (\hat{\bm v}_{\rm L}\cdot \tilde{\bm \nabla} - \tfrac{\imu }{2}\tilde \varphi) \psi
= \left[
-1 + |\psi|^2 - \tilde {\bm \nabla}^2 + \imu k (\tilde \partial_x^3 
- 3\tilde \partial_x\tilde \partial_y^2)
\right]\psi
, \\
&\left(\frac{1}{u} \tilde{\bm \nabla}^2 - |\psi|^2 \right) \tilde \varphi 
= - \imu \hat{\bm v}_{\rm L}\cdot (\psi \tilde {\bm \nabla} \psi^* - \psi^*\tilde {\bm \nabla} \psi)
, \\
&v=\frac{\Gamma v_{\rm L}}{|\al|\xi}
,\ \ 
k=\frac{K}{|\al|\xi^3}
,\ \ 
\psi = \frac{\Delta}{|\Delta_\infty|}
,\ \ 
\tilde \varphi = \frac{\varphi}{v_{\rm L} / 4e\xi}
,
\end{align}
where $|\Delta_\infty| = \sqrt{|\al|/\beta}$ is the gap function for a uniform bulk and 
$\xi=\sqrt{\gm/|\al|}$ is the coherence length.
We have also defined the unit vector $\hat{\bm v}_{\rm L} = \bm v_{\rm L}/v_{\rm L}$ and the dimensionless derivative 
$\tilde {\bm \nabla} = \xi \bm \nabla$.
In the equation for the scalar potential, we have introduced the temperature independent parameter 
$u = \xi^2/\ell_E^2$,
where the length $\ell_E$ is the electric-field penetration depth 
given by
\begin{align}
\ell_E &= \sqrt{\frac{\sg_n}{8e^2\Gamma|\Delta_\infty|^2}}
. \label{eq:def_ell_E}
\end{align}
For an ordinary metal, the parameter $u$ is an order of unity constant \cite{Kopnin_book}.

When we use the relation $\Gamma=\pi\nu \hbar / 8k_{\rm B} T_0$ with the density of states 
$\nu = m/2\pi\hbar^2$ for a two-dimensional electron gas,
where we have restored $\hbar$ and $k_{\rm B}$,
 the parameter can be written as
\begin{align}
u = \frac \pi 3 \cdot \frac{e^2/h}{\sg_n}\cdot \frac{E_{\rm F}}{k_{\rm B}T_0}
,
\end{align}
where $h/e^2 = 25813 \,[\Omega]$ is the von Klitzing constant.
For usual BCS superconductors, the ratio between $e^2/h$ and the normal conductivity $\sg_n$ is 
comparable to $k_BT_0/E_{\rm F}$.
For example, in the monolayer MoS$_2$ \cite{Wakatsuki}, we have 
$1/\sg_n = 140 \,[\Omega]$, $T_0 = 8.8 \,[{\rm K}]$, $E_{\rm F}=150 \,[{\rm meV}]$, 
and obtain $u\simeq 1.2$.

The energy dissipation is also derived from the time-derivative of free energy
\begin{align}
\partial_t F_s &= \int \diff \bm r\left[
-2\Gamma  \left| \left( \partial_t + 2\imu e\varphi \right)\Delta  \right|^2
+ \varphi \bm \nabla \cdot \bm j_s
\right]
.
\end{align}
Using the electromagnetic energy conservation law $\partial_t F_n = - \int \diff \bm r\, \bm j\cdot \bm E$ derived from the Maxwell equation, we obtain the equation of continuity
for the energy density:
\begin{align}
&\partial_t F + \int \diff \bm r \bm \nabla\cdot \bm j_F = - \int \diff \bm r w
, \\
&\bm j_F = - \varphi \bm j_s
, \\
&w = 2\Gamma  \left| \left( \partial_t + 2\imu e\varphi \right)\Delta  \right|^2
+ \sg_n \bm E^2
,\label{eq:dissipation}
\end{align}
where $\bm j_F(\bm r,t)$ is an energy current density and $w(\bm r,t)$ is a dissipation rate.
Thus we have two kinds of the dissipation terms originating from $\Gamma$ and $\sg_n$.

\subsubsection{Perturbative analysis}
Let us now analyze the differential equation perturbatively with respect to $K$ and $v_{\rm L}$.
We expand the physical quantity $A$ in general as 
\begin{align}
A=A_0 + kA_k + vA_v+kvA_{kv} + O(k^2,v^2)
.
\end{align}
For $O(1)$, we obtain
\begin{align}
&0
= \left(
-1 + |\psi_0|^2 - \tilde {\bm \nabla}^2 
\right) \psi_0 
,\\
&\left(\frac{1}{u}\tilde {\bm \nabla}^2 - |\psi_0|^2 \right) \tilde \varphi_0 = - 2\imag
\hat{\bm v}_{\rm L}\cdot \psi_0 \tilde {\bm \nabla} \psi_0^*
,
\end{align}
and for $O(K)$
\begin{align}
&0
= 
-\psi_k + 2\psi_k |\psi_0|^2+\psi_0^2\psi_k^* - \tilde {\bm \nabla}^2 \psi_k 
+ \imu (\tilde \partial_x^3 - 3\tilde \partial_x \tilde \partial_y^2)\psi_0
,\\
&\frac{1}{u} \tilde{\bm \nabla}^2 \tilde \varphi_k
- |\psi_0|^2  \tilde \varphi_k 
- 2\real(\psi_0\psi_k^*)\tilde \varphi_0
\nonumber \\
&= 2\imag
 \hat{\bm v}_{\rm L}\cdot (\psi_0 \tilde {\bm \nabla} \psi_k^* + \psi_k \tilde {\bm \nabla} \psi_0^*)
.
\end{align}
For $O(v_{\rm L})$, the equations are
\begin{align}
&(\hat{\bm v}_{\rm L}\cdot \tilde{\bm \nabla} - \tfrac{\imu}{2} \tilde \varphi_0) \psi_0
= 
-\psi_v + 2\psi_v |\psi_0|^2+\psi_0^2\psi_v^* - \tilde{\bm \nabla}^2 \psi_v
, \\
&\frac{1}{u}\tilde {\bm \nabla}^2 \tilde \varphi_v
- |\psi_0|^2  \tilde \varphi_v
- 2\real(\psi_0\psi_v^*)\tilde \varphi_0
\nonumber \\
&= 2\imag \hat{\bm v}\cdot (\psi_0 \tilde {\bm \nabla} \psi_v^* + \psi_v \tilde {\bm \nabla} 
\psi_0^*)
.
\end{align}
Finally for $O(Kv_{\rm L})$
\begin{align}
&\hat{\bm v}_{\rm L}\cdot \tilde {\bm \nabla} \psi_{k} - \tfrac{\imu}{2} (\tilde \varphi_0 \psi_k
+ \tilde \varphi_k \psi_0)
= 
-\psi_{kv} + \psi_0^2\psi_{kv}^* 
\nonumber \\
&
+ 2 \psi_{kv} |\psi_0|^2
+ 2 \psi_k \psi_0\psi_v^*
+ 2 \psi_v \psi_0 \psi_k^*
+ 2 \psi_k \psi_v \psi_0^*
\nonumber \\
&- \tilde {\bm \nabla}^2\psi_{kv} + \imu(\tilde \partial_x^3 - 3\tilde \partial_x 
\tilde \partial_y^2)\psi_{v}
,
\end{align}
and 
\begin{align}
&\frac{1}{u}\tilde {\bm \nabla}^2 \tilde \varphi_{kv}
- |\psi_{0}|^2\tilde \varphi_{kv}
\nonumber \\
&
- 2\real(\psi_{0} \psi_{kv}^* \tilde \varphi_{0}
+ \psi_{0} \psi_{k}^* \tilde \varphi_{v}
+ \psi_{0} \psi_{v}^* \tilde \varphi_{k}
+ \psi_{k} \psi_{v}^* \tilde \varphi_{0})
\nonumber \\
&= 2\imag \hat{\bm v}_{\rm L}\cdot 
(\psi_0 \tilde {\bm \nabla} \psi_{kv}^* + \psi_{kv} \tilde {\bm \nabla} \psi_0^* + \psi_k
\tilde {\bm \nabla} \psi_v^* + \psi_v \tilde {\bm \nabla} \psi_k^*)
.
\end{align}
The force balance relation becomes
\begin{align}
\frac{\bm j_{\rm unif} \times \bm \Phi_0}
{\Gamma v_{\rm L}|\Delta_\infty|^2} 
&= 
\left[
f_1 
+
kvf_2
\begin{pmatrix}
\hat v_{\mrm Ly} & \hat v_{\mrm Lx} \\
\hat v_{\mrm Lx} & -\hat v_{\mrm Ly}
\end{pmatrix}
\right]
\hat{\bm v}_{\rm L}
+O(k^2,v^2)
, \label{eq:force_balance_TDGL}
\end{align}
where $f_1$ and $f_2$ are linear and nonlinear coefficients in the force-velocity relation.
The coefficient $f_1$ is a sum of the contributions from $u$-independent Tinkham mechanism 
(time-dependence of amplitude of the gap) and $u$-dependent Bardeen-Stephen mechanism 
\cite{Tinkham_book,Kopnin_book}.
Here the transverse friction force appears in addition to the longitudinal one.
This form can be derived by a symmetry consideration, and have also been checked numerically.

The energy dissipation part can also be expanded with respect to $k$ and $v$.
In the dimensionless form, we can write it as
\begin{align}
\tilde w =
\frac{w}{\sg_n \varphi_0^2 / \xi^2}
= 4u 
\left|\left(
\hat{\bm v}_{\rm L}\cdot \tilde{\bm \nabla} +\tfrac{\imu}{2}\tilde \varphi
\right) \psi 
\right|^2
+ \left|
\tilde {\bm \nabla}\tilde \varphi 
\right|^2
.
\end{align}
The $O(v_{\rm L}K)$ component for $\tilde w$, or equivalently $O(v_{\rm L}^3K)$ for $w$, is responsible 
for the dissipation from nonreciprocal vortex dynamics.

Let us look at the spatial dependence of the above physical quantities.
We use the two-dimensional polar coordinate $\bm r = (r,\phi)$ and the zeroth-order solution 
can be written as $\psi_0(\bm r) = f(r)\epn^{\imu\phi}$ for an isolated vortex.
From the differential equation for $\psi_0$, we find the asymptotic behavior $f\propto r$ 
for $r\rightarrow 0$ and $f\rightarrow 1$ for $r\rightarrow \infty$ \cite{Kopnin_book}.
Among functional forms that satisfy these limiting behaviors, we choose $f(r) = \tanh(a r/\xi)$ 
and the constant is determined as $a=\sqrt{\tfrac{3}{8}}$ by using the differential equation at 
small $r$.
Then the physical quantities such as $\psi_{0,k,v,kv}$, $\tilde \varphi_{0,k,v,kv}$ and 
$\tilde w_{0,k,v,kv}$ are  numerically calculated by solving the linear differential equations 
derived above.

\begin{figure*}[t]
\begin{center}
\includegraphics[width=160mm]{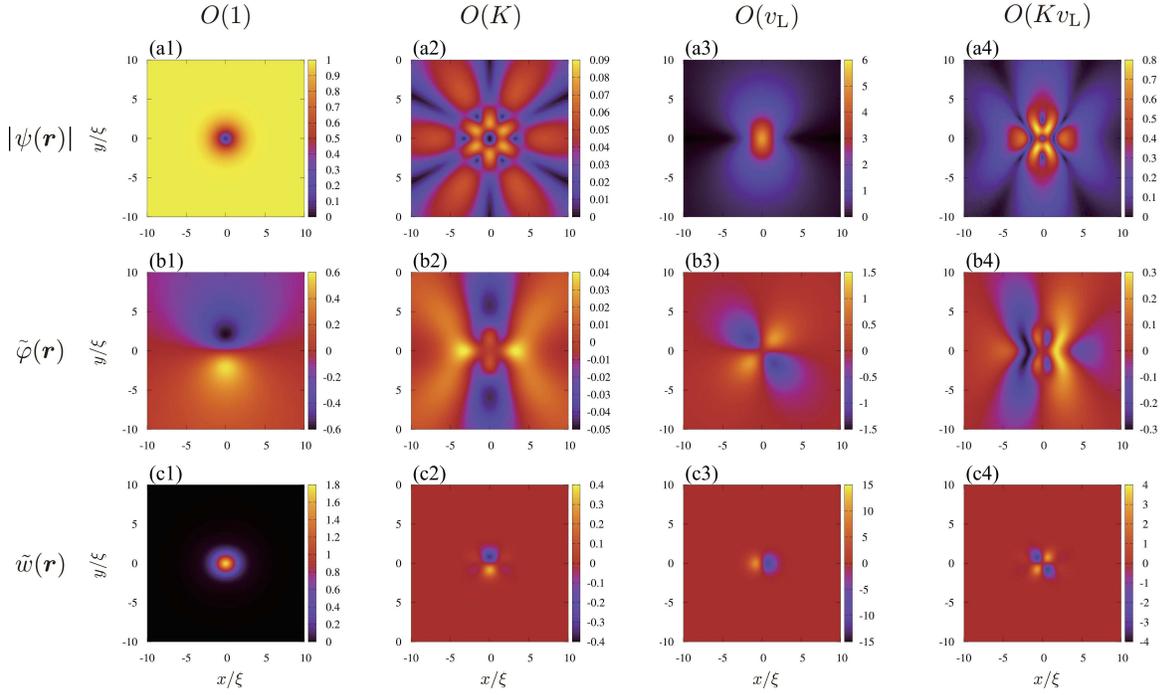}
\caption{
Spatial dependences of the amplitude of wave functions, scalar potential, and energy 
dissipation rate:
(a1) $|\psi_0|$, 
(a2) $|\psi_k|$, 
(a3) $|\psi_v|$, 
(a4) $|\psi_{kv}|$;
(b1) $\tilde \varphi_0$, 
(b2) $\tilde \varphi_k$, 
(b3) $\tilde \varphi_v$, 
(b4) $\tilde \varphi_{kv}$;
(c1) $\tilde w_0$, 
(c2) $\tilde w_k$, 
(c3) $\tilde w_v$, 
(c4) $\tilde w_{kv}$.
We have chosen $\bm v_{\rm L} = v_{\rm L}\hat{\bm x}$ and used $u=1$ (i.e. $\ell_E = \xi$).
The system size is $L\times L$ with $L=40\xi$ and the number of mesh is $N_L\times N_L$ 
with $N_L=300$.
}
\label{fig:map}
\end{center}
\end{figure*}

Figures~\ref{fig:map}(a1--4) shows the spatial dependences of the amplitudes of gap functions.
The originally circular shape in Fig.~\ref{fig:map}(a1) is modified by $K$ and $v_{\rm L}$, and six-fold 
and two-fold patterns appear in Figs.~\ref{fig:map}(a2) and (a3), respectively.
The more complex pattern is seen in the higher-order contribution $\psi_{kv}$ 
[Fig.~\ref{fig:map}(a4)].
The scalar potentials are also shown in Figs.~\ref{fig:map}(b1--4).
The dipolar field is generated in $\varphi_0$ which causes the electric field inside the normal core.
We note that the dimensionless $\tilde \varphi_0$ is $O(1)$ but the scalar potential $\varphi_0$ 
is $O(v_{\rm L})$ contribution.
Then $\varphi_v$ shown in Fig.~\ref{fig:map}(b3) is an $O(v_{\rm L}^2)$ contribution, having 
the quadrupolar distribution. 
The situation is further modified in the presence of the cubic term as in Figs.~\ref{fig:map}(a2) 
and (a4).
We show in Figs.~\ref{fig:map}(c1--4) the energy dissipation rate as a function of spatial coordinates.
The dissipation occurs in the region with $r\lesssim\ell_E$ as shown in Fig.~\ref{fig:map}(c1) 
(note that $W_0$ is an $O(v_{\rm L}^2)$ contribution).
This also applies at higher orders shown in Figs.~\ref{fig:map}(c2--4), where the spatial anisotropy 
is introduced.

\begin{figure}[t]
\begin{center}
\includegraphics[width=70mm]{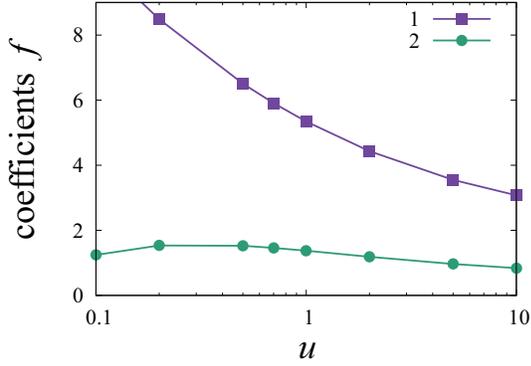}
\caption{
Parameter $u$ dependence of the friction force coefficients defined in 
Eq.~\eqref{eq:force_balance_TDGL}.
The system size and number of spatial mesh are same as Fig.~\ref{fig:map}.
}
\label{fig:u_depend}
\end{center}
\end{figure}

We are also interested in the parameter $u=\xi^2/\ell_E^2$ dependence of the physical quantities.
Figure~\ref{fig:u_depend} shows the force coefficients $f_1$ and $f_2$ as a function of $u$.
For the larger $u$, meaning shorter electric-field penetration length, the value of $f_1$
becomes 
smaller.
This is because the region where the dissipation occurs shrinks to make a weaker friction force.
The value of $f_2$ at sufficiently large $u$ also becomes small due to the same reason as $f_1$.
On the other hand, $f_2$ at small $u$ decreases in contrast with $f_1$.
While the reason for the behavior is not easily understood in an intuitive way, in any case we have here confirmed that the coefficients of $f_1$ and $f_2$ are order of unity in the experimentally relevant regime with $u \sim 1$.

\subsubsection{Flux-flow conductivity}

Using the relation
$\bm E_0 = \bm B \times \bm v_{\rm L}$ for an electric field generated by the motion of the magnetic flux, 
which is derived from the Josephson relation,
we obtain the electrical conductivity as
\begin{align}
\bm j_{{\rm unif}} &= \frac{\Gamma |\Delta_\infty|^2}{\Phi_0 B_z}
\left[
f_1 + \frac{f_2 \Gamma K}{|\al|^2\xi^4B_z}
\begin{pmatrix}
 E_{0x} & -E_{0y} \\
-E_{0y} & -E_{0x}
\end{pmatrix}
\right] \bm E_0
\\
&= \sg_{1\mrm v} \bm E_0 + \sg_{2\mrm v}
\bm F(\bm E_0)
,
\end{align}
where the vector $\bm F$ is same as that in Eq.~\eqref{eq:TMD_current}.
To express the flux-flow conductivity in terms of normal conductivity, we use the electric field 
penetration depth $\ell_E$ and the upper critical field given by $B_{c2} = \Phi_0/2\pi\xi^2$.
Then the conductivities are given by
\begin{align}
\sg_{1\mrm v} &= \frac{u f_1(u)}{4\pi} \sg_n \frac{B_{c2}}{B_z}
, \\
\sg_{2\mrm v} &= \frac{u f_2(u)}{4\pi} \sg_n \frac{B_{c2}\Gamma K}{\gm^2 B_z^2}
.
\end{align}
The $\gamma$-value from nonreciprocal vortex dynamics is given by
\begin{align}
\gm_{\rm S}^{\mrm v} &= \frac{\sg_{2\mrm v}}{\sg_{1\mrm v}^2B_zW} =
 \frac{4\pi f_2(u)}{u f_1(u)^2} \cdot
 \frac{\Gamma K}{\gm^2\sg_n B_{c2}B_z W }
.
\end{align}
This expression does not depend on $B_z$ since $K$ is proportional to $B_z$ due to the Zeeman effect.
We note that the parameters $\sg_n$, $\gm$ and $K$ are in general dependent on the purity of 
the sample (see Appendix B).

Let us compare this result with the $\gm$-value ($\gm_{\rm S}^{\mrm f}$) from superconducting 
fluctuation for $T\gtrsim T_0$ \cite{Wakatsuki}.
Noting $u\sim 1$, the ratio is estimated by
\begin{align}
\frac{\gm_{\rm S}^{\mrm v}}{\gm_{\rm S}^{\mrm f}} &\sim
\frac{T_0^2}{E_{\rm F}(T_0-T)}
.
\end{align}
Thus we obtain the same conclusion as the extended Bardeen-Stephen approach in Sec.~III\,A\,2.

\subsection{Ratchet motion of vortex} \label{Subsec:Ratchet}

As discussed above, for $T<T_0$ and in the presence of the finite out of plane magnetic field, the motion of 
quantum vortices penetrating the superconductor gives a dominant contribution to the voltage drop.
Here we consider another effect for vortex dynamics.
In the superconductor without inversion center, vortices driven by the external charge current feel an 
asymmetric pinning potential acting on them (ratchet effect). 
This effect has been proposed to control the vortices in superconductors \cite{Lee99, Villegas03, Villegas05}.
New perspective here is that the vortex ratchet effect {\it naturally} appears as a consequence of disorders in noncentrosymmetric system, which is distinct from the previously discussed  {\it artificially} developed inversion-broken environment.
In this section, we analyze the classical motion of 
vortices under the asymmetric periodic potential and discuss the nonreciprocal transport there.
The relevant parameters such as potential height and its spatial 
periodicity are estimated 
from pinning properties such as a magnitude of the critical current density where the vortices are depinned.
The relation to the recent experimental results \cite{Wakatsuki,Saito_private} will also be discussed in Sec.~IV\,C\,2.

The classical equation of motion of the vortices in the diffusive limit is modeled as
\begin{equation}
\eta \frac{\mathrm{d} x}{\mathrm{d} t} = -\frac{\partial U}{\partial x } + F - \sqrt{2\eta T}\xi(t)
,
\end{equation}
with $U$ being the pinning potential, which we here assume to be periodic for simplicity; 
$U(x+L)=U(x)$, and $F$ is the uniform force. The last term is the Langevin force with zero mean 
and unit dispersion (Gaussian white noise). $\eta$ is the viscous friction coefficient, and $T$ is the 
temperature.
By solving the corresponding Fokker-Planck equation
\begin{equation}
\frac{\partial p}{\partial t}= \frac{1}{\eta} \frac{\partial }{\partial x}
\left[ \left( \frac{\partial U}{\partial x} -F \right) p + T \frac{\partial p}{\partial x}\right]
,
\end{equation}
for the distribution function $p(x,t)$, we obtain the steady velocity of the vortex after long time expressed 
as \cite{Reimann}
\begin{equation}
v_{\rm L}=\frac{L}{\beta\eta} \frac{1-e^{-\beta F L}}{ \int_0^L d y \ I_0(y) e^{ -\beta Fy} }
,
\end{equation}
where we have introduced 
\begin{equation}
I_0(y) = \int_{x_0}^{x_0+L} d x \ e^{\beta \left[ U(x)-U(x-y)\right]}
,
\end{equation}
with $\beta=1/T$ being the inverse temperature. We can choose $x_0$ arbitrarily due to the periodicity 
of the potential. Potential is called symmetric when there exists certain choice of 
$x_0 \in \left[0,L\right)$ such that $U(x)=U(x_0-x)$ is satisfied. Here, we can choose $x_0=0$ 
by shifting the $x$ coordinate. By using the evenness and the periodicity of the potential, we can 
easily prove $v_{\rm L}(-F)=-v_{\rm L}(F)$ which show the absence of the nonreciprocal transport. 

For asymmetric potentials, the situation is different. The velocity $v$ is expanded with respect to 
the force $F$ as
\begin{align}
v_{\rm L}&= q_1 F + q_2 F^2 +O\left(F^3\right)  
,
\end{align}
where the coefficients are
\begin{align}
q_1&=\frac{L}{\beta\eta} \frac{\beta L}{  \int_0^L d y \ I_0(y) } 
,\\
q_2&=\frac{L}{\beta\eta} \frac{\beta ^2 L  \int_0^L d y 
\left(y-\frac{L}{2}\right) I_0(y) }{\left[ \int_0^L d y \ I_0(y) \right]^2}
,
\end{align} 
where $q_2$, nothing but a hallmark of the nonreciprocal transport, survives only for 
asymmetric potentials as discussed above. 

Hereafter, we adopt the fully asymmetric periodic potential for simplicity
\begin{equation}
U(x) = U_0 \ \mathrm{saw} \left( \frac{x}{L}\right) = U_0 \frac{x}{L} 
\quad ( 
\mathrm{mod} \ L
)
, \label{eq:saw}
\end{equation}
by using so-called sawtooth function. $U_0 (>0)$ is the height of the pinning potential.
In this potential, the steady velocity is calculated as
\begin{widetext}
\begin{align}
v_{\rm L} = \frac{1}{\beta \eta L }
\frac{\beta^2  (U_0-FL)^2}{\beta  FL - \beta U_0 -\sinh ( \beta U_0  )
+(\cosh ( \beta U_0  )-1) \coth \left(\frac{\beta  FL}{2}\right) } = q_1F + 
q_2F^2+O\left(F^3\right)
,
\end{align}
\end{widetext}
with the response coefficients
\begin{align}
  q_1 &= \frac{1}{2 \eta}\frac{\beta ^2  U_0^2 }{ \cosh (\beta U_0 )-1} =
\begin{cases}
\displaystyle
    \frac{1}{\eta} -\frac{\beta^2 U_0^2}{12\eta}& (\beta U_0 \rightarrow 0) 
\\[4mm]
\displaystyle
    \frac{\beta^2 U_0^2}{\eta}e^{-\beta U_0}& (\beta U_0 \rightarrow \infty)
  \end{cases}
,
\end{align}
and
\begin{align}
  q_2 &=\frac{L\beta}{4\eta}\frac{ \beta^3 U_0^3 +\beta^2 U_0^2 \sinh (\beta U_0  )
+4 \beta U_0  -4 \beta U_0    \cosh (\beta U_0 )}{ (\cosh (\beta U_0 )-1)^2}\nonumber \\
&=\begin{cases}
\displaystyle
    \frac{L}{\eta}\frac{\beta^4 U_0^3}{360}& (\beta U_0 \rightarrow 0) 
\\[4mm]
\displaystyle
    \frac{L}{\eta}\frac{\beta^3 U_0^2}{2}e^{-\beta U_0}& (\beta U_0 \rightarrow \infty).
  \end{cases}
.
\end{align}
Since the voltage drop originating from the motion of vortices is $V=B_z L_x v_{\rm L}$ and the force 
acting on vortices is $F=\frac{\phi_0}{W} I$, we can calculate the coefficients in current-voltage relation [Eq.(\ref{Eq:IV})]
as $a_1(B_z,T) = \phi_0\frac{L_x}{W} B_z q_1(T) $ and 
$a_2(B_z,T) = \phi_0^2\frac{L_x}{W^2} B_z q_2(T)$, both of which are proportional 
to $B_z$. The nonreciprocal $\gamma '$ parameter defined as $R=R_0(1+\gamma' I)$ with $V=RI$ is 
expressed as
\begin{align}
\gamma' &=\frac{q_2}{q_1} \frac{ \phi_0}{W}= \frac{ \phi_0 L }{W} 
\frac{  U_0^2 \beta ^2+  \beta U_0  \sinh ( \beta U_0 )-4 \cosh ( \beta U_0 )+4 }{4 U_0 
\sinh ^2\left(\frac{ \beta U_0 }{2}\right)}\nonumber \\
&=\begin{cases}
\displaystyle
   \frac{\phi_0 L}{W} \frac{\beta^4 U_0^3}{360} & (\beta U_0 \rightarrow 0) 
\\[4mm]
\displaystyle
   \frac{\phi_0 L}{W} \frac{\beta}{2}& (\beta U_0 \rightarrow \infty)
  \end{cases}
, \label{eq:gamma_ratchet}
\end{align}
which monotonically decrease as raising temperature. 
Note that the exponential temperature dependence vanishes for the $\gm$-value.

In this calculation, we have neglected the vortex-vortex interaction. 
This assumption is justified 
for small magnetic field where the vortices are dilute enough.
We also note that the ratchet effect is active for trigonal symmetry, but is not relevant for the systems with $C_\infty$ and hexagonal symmetries where no ratchet potential is present.
In the latter cases, the effect from asymmetric spin-orbit coupling plays the more dominant role.

\section{Discussion}

Here we discuss expected noreciprocal charge signals in 
2DNS.
Table \ref{fig:table} summarizes the 
nonreciprocal $I$-$V$ characteristics
 in Rashba superconductors, TI surface and TMD for both above and below transition temperature.
The magnetic fields are applied parallel to the two-dimensional plane for Rashba and TI based systems, and applied perpendicular to the layer for TMD.
While the other configurations can in principle be possible, the information of this paper gives a firm basis to explore the further properties.
Below we discuss each system separately.

\subsection{Rashba superconductors}

Here we consider the typical $\gm$-value for Rashba superconductors.
With electron or hole doping, the Fermi energy can be tuned and the behavior is dependent on the sign of $E_{\rm F}$.
Let us begin with the $E_{\rm F}<0$ case.
In this case, the normal contribution to $\gm$-value becomes finite as shown in Ref.~\onlinecite{IdeueHamamoto} (see also Table~\ref{fig:table}).
The typical values for BiTeBr have been estimated \cite{IdeueHamamoto}
 by using the effective mass h$m = 0.15m_e$, the Rashba parameter $\alpha = 2.00$ eV\,\AA, and the g-factor $g = 60$.
In the normal state with $E_{\rm F} = -0.01$ eV,
the amplitude of the magnetochiral 
anisotropy is estimated as $W\gm_{\rm N} \simeq 2 \times 10^{-5}$ ${\rm T^{-1}A^{-1}m}$.

The system crossovers into the paraconductivity region with approaching to the mean-field transition temperature $T_0$ from above.
Here, as discussed in Sec.~\ref{sec:rashba_sc}, the parity mixing contribution becomes irrelevant and the cubic term in GL free energy instead becomes dominant.
The paraconductivity is then given by Eq.~\eqref{Eq:gammacubic}.
The ratio between normal and superconducting states is given by $\gm_{\rm S}/\gm_{\rm N} \sim E_{\rm F}/T_0$, and thus the nonreciprocal signal is strongly enhanced by the appearance of a small energy scale $T_0$ for superconductors.

The fluctuation contribution above $T_0$ further crossovers to vortex contribution at lower temperatures than $T_0$.
Below $T_0$, the pair amplitude sufficiently develops and the 
free vortices,
which are generated thermally above KT transition temperature in the present system, start to play an important role for transport phenomena.
For $E_{\rm F}<0$ case, the cubic term effectively renormalizes superfluid density under the transport current as discussed in Sec.~\ref{sec:renormalization}.
As a result, the friction force and KT transition temperature are modified and have different values depending on the direction of source currents.
The former causes the characteristic temperature dependence in the $\gm$-value as $\gm_{\rm S} \propto (T_0-T)^{-1}$ near $T_0$ $(>T_{\rm KT})$.
Note that this expression smoothly connects to the fluctuation contribution for $T>T_0$, and does not show divergence in reality.
On the other hand, the modification of the KT transition temperature shows the divergent $\gm$-value as $\gm_{\rm S}\propto (T-T_{\rm KT})^{-3/2}$ near the KT transition point.
For $T<T_{\rm KT}$, vortex and antivortex are bound, and the linear transport coefficient finally vanishes.
The third-order term with $a_3$ then becomes the relevant one in the current-voltage relation.
The nonreciprocity is reflected in the higher order term with $a_4$ in this case.
More detailed investigation of these higher-order contributions remains to be clarified in the future.

Now, we switch our focus to the $E_{\rm F}>0$ case.
Although the normal state contribution to $\gm$-value is absent in this situation \cite{IdeueHamamoto},
the paraconductivity is finite.
There are two contributions to paraconductivity: one from parity mixing and the other from $\bm q$-cubic term in the GL theory.
Here, as the ratio is calculated in Eq.~\eqref{Eq:pmcratio}, the parity-mixing contribution in Eq.\eqref{Eq:gammapm} is much larger than the other.
The $\gm$-value in this case has been estimated in Ref.~\cite{WakatsukiNagaosa} for BiTeBr with superconducting proximity effects.
On the othe hand, the~\LAOSTO~interface \cite{Reyren, Caviglia, Maniv, Herranz} is also a typical two-dimensional Rashba superconductor.
Its carrier density is given by $n \sim 10^{13} \mathrm{cm}^{-2}$, spin--orbit field is $B_\mathrm{SO} = \frac{m^2\al^2}{|e|} \sim 1 \mathrm{T}$, the Debye temperature is $T_\mathrm{D} \sim 400 \mathrm{K}$, and the mean-field transition temperature is $T_0 \sim 100 \mathrm{mK}$.
If we assume $r_\mathrm{t} = 0.1$ and the typical sample width $W = 10^{-6}$m, the $\gamma$-value is estimated as $\gamma_\mathrm{S} \sim 8 \times 10^{4} \mathrm{T^{-1}A^{-1}}$, which is much larger than the previous studies \cite{Rikken1, Rikken2, Pop, Krstic, Rikken3}.

At lower temperature below $T_0$, the vortex contribution dominates over the paraconductivities.
For $E_{\rm F}>0$ case, the two-component gap parameter need to be considered for vortex dynamics.
While detailed studies remain unexplored, the vortex contribution should be present and is expected to cause a singular behavior around the KT transition point as in the $E_{\rm F}<0$ case.

\subsection{TI surface}

We here discuss the surface of topological insulators plus superconducting proximity effect.
The Hamiltonian has the $\bm k$-linear term, but the nonreciprocal charge transport is absent with this term only.
We have thus considered the two more terms to generate the nonreciprocity: parabolic term and hexagonal warping term.
Let us first consider the contribution from parabolic term based on Eq.\eqref{eq:TI_1}.
The functional forms of the $\gm$-values are listed in Table~\ref{fig:table}.
The ratio between normal and superconducting states is given by $\gm_{\rm S}/\gm_{\rm N} \sim E_{\rm F}/T_0$.
Hence, the nonreciprocal transport signal is enhanced in superconducting state.
To estimate the typical value,
we use the expression in Eq.~\eqref{eq:TI_1} with $\hbar$, $\kB$, and $\muB$ recovered.
If we assume $v_{\rm F}=2.84$ eV{\AA} and $\frac{1}{2m}=41.1$ eV{\AA}$^2$ in 
Bi$_2$Te$_3$ \cite{Kim}, and $\EF = 0.1$ eV, $T_0=10$ K, and $W=100$ $\mu$m as typical values, 
we obtain $\gamma_{\mathrm S} \approx 0.33$ $\mathrm{A}^{-1} \mathrm{T}^{-1}$.
On the other hand, the contribution from the hexagonal warping is given in Eq.~\eqref{eq:TI_2}, and is here also estimated.
Assuming the same parameters above and $\sqrt{\lambda}=250$ eV{\AA}$^3$ in 
Bi$_2$Te$_3$ \cite{Kim, Fu}, 
we obtain $\gamma_{\mathrm S} \approx 0.11$ $\mathrm{A}^{-1} \mathrm{T}^{-1}$.
Therefore, the amplitude is comparable to that by the parabolic term.

At lower temperatures, the vortex contribution becomes dominant as in the Rashba superconductors.
Since the magnetic field is applied along the two-dimensional plane and there is the cubic term in GL free energy, the behavior is essentially same as the Rashba superconductors with $E_{\rm F}<0$.
Namely, the thermally generated vortex in the $T_{\rm KT} < T \lesssim T_0$ region creates the characteristic magnetochiral anisotropy in the forms 
$\gm_{\rm S} \propto (T_0-T)^{-1}$ for $T\rightarrow T_{0}$
and
$\gm_{\rm S} \propto (T-T_{\rm KT})^{-3/2}$ for $T\rightarrow T_{\rm KT}$.
We note that the above transport coefficients are written in the form $V= a_1 I(1+\gm B I)$, i.e. the magnetic field $B$ enters only with $\gm$-value.
As shown below, however, the situation can qualitatively change if the magnetic field is applied perpendicular to the plane.

\subsection{TMD}

\subsubsection{Paraconductivity and intrinsic vortex-flow contribution}
We estimate the physical quantities of the clean MoS$_2$.
Let us begin with the normal contribution well above $T_0$.
This system has a valley degrees of freedom, whose contribution to the normal $\gm$-value per valley is listed in Table~\ref{fig:table}.
With the situation in MoS$_2$, however, the $\gm$-value for each valley has different sign and vanishes if we sum up both the contributions \cite{Wakatsuki}.
Near $T_0$, the paraconductivity contribution is developed as given in Eq.~\eqref{eq:gamma_TMD}.
Using $2m\lambda/\hbar^2=-0.49\AA$ and $\Delta_{\rm SO}\simeq 7.5$ meV,
and $T_0 = 8.8$ K for monolayer MoS$_2$,
the $\gm$-value 
from the superconducting fluctuation reaches $\gm_{\rm S} \simeq 250 {\rm T^{-1}A^{-1}}$ for the sample width $W \simeq 3 \mu {\rm m}$, as shown in Ref.~\onlinecite{Wakatsuki}.

Below $T_0$, here again the vortex contribution becomes relevant for the $\gm$-value, but the situation is different from the Rashba and TI based systems with KT transition.
Namely, the vortices with the same vorticity are induced by the out-of-plane magnetic field and the KT transition is washed away for $B\neq 0$.
As a result, the number of vortices are determined by the external field $B$, and the ordinary resistivity $a_1(B)$ is proportional to $B$.
We reflect this situation by denoting $I$-$V$ characteristic as $V=a_1' B I + a_2(B) I^2$.
As for the coefficient $a_2(B)$, there are two types of contributions. 
One is from the $\bm q$-cubic term in the GL free energy and $a_2$ is proportional to $B^2$.
The other is from the ratchet potential for vortices, and $a_2\propto B$ is satisfied.
Thus the $B$ dependence of $a_2$ clearly distinguishes the underlying mechanism to generate nonreciprocal charge transport.
Since the latter effect is discussed in detail in the next subsection, we here focus on the cubic-term contribution.
The $I$-$V$ relation is then written as $V=a_1'BI(1+\gm_{\rm S}BI)$.

As discussed in Sec.~\ref{sec:renormalization}, the $\bm q$-cubic term in GL free energy effectively renormalizes the superfluid density.
We have applied the generalized Bardeen-Stephen approach to this system: the force balance between driving (Magnus) force and viscous force acting on the vortex is considered.
The nonreciprocity enters in the 
friction force through the anisotropically renormalized superfluid density.
We have then found that the $\gm$-value from vortex dynamics is given by $\gm_{\rm {S}}^{\rm v} = \gm_{\rm {S}}^{\rm f} \times \frac{T_0^2}{E_{\rm F}(T-T_0)}$ with $\gm_{\rm S}^{\rm f}$ being the superconducting fluctuation contribution, which has a large value near the mean-field transition temperature $T_0$.
Here the temperature dependence enters through the unrenormalized superfluid density which behaves as $n_s\propto (T_0 - T)$.
Such behavior has further been justified by the TDGL approach as demonstrated in Sec.~III\,B.
If the results are extrapolated to zero temperature, we get $\gm^{\rm v}_{\rm {S}}/\gm^{\rm f}_{\rm {S}} = T_0 / E_{\rm F}$.
Since the magnitude of the paraconductivity contribution is roughly given by $\gm_{\rm S}^{\rm f}\sim 10^2 {\rm T^{-1}A^{-1}}$ as discussed above, the order of magnitude for the vortex contribution is $\gm_{\rm S}^{\rm v} \sim 1 {\rm T^{-1}A^{-1}}$ in the low-temperature limit.
This is much smaller than the observed values at low $T$ in the monolayer MoS$_2$ \cite{Wakatsuki}, and hence we need another mechanism to account for the experimental results.

\subsubsection{Ratchet effect of vortex flow}

We now consider the nonreciprocal transport from ratchet effect of vortex dynamics based on Eq.~\eqref{eq:gamma_ratchet}.
Phenomenological parameters such as periodicity $L$ of the pinning potential, friction coefficient $\eta$, and potential height $U_0$ are estimated for MoS$_2$ using 
experimental data.
The parameter $L$
is determined from the mean distance of 
pinning centers. This can be estimated from pinning-depinning transition point in the 
magnetoresistance measurement, that is about 
$B_z\simeq 0.2\mathrm{T}$ \cite{Saito_private}
At this transition point, all the pinning centers are assumed to be filled with vortices. 
The total flux is $B_z L_x W = N_{\rm v} \Phi_0$ then the vortex number density is 
$n_{\rm v}=N_{\rm v}/(L_x W) = B_z/\Phi_0$. Thus the mean distance between vortices is
\begin{equation}
L\sim \frac{1}{\sqrt{n_{\rm v}}} = \sqrt{\frac{\Phi_0}{B_z}} \sim 10^{-7} \mathrm{m} .
\end{equation}

The parameter $\eta$ is estimated by the normal state resistivity. In the absence of the 
pinning potentials, $(U_0=0)$, the $I$-$V$ characteristic becomes $v_{\rm L}=F/\eta$ or
\begin{equation}
V = B_z v_{\rm L} L_x = B_z \frac{F}{\eta} L_x = B_z \frac{\Phi_0}{\eta} \frac{L_x}{W} I = R I
,
\end{equation}
in the ohmic region. The resistivity $R=B_z \frac{\Phi_0}{\eta} \frac{L_x}{W}$ should 
be same as the normal state resistivity $R_{n}$ when $B_z=
B_{\rm c2}
\simeq 0.1\ \mathrm{T}$.
Thus, the parameter $\eta$ is estimated as
\begin{equation}
\eta=\frac{\Phi_0 
B_{\rm c2}}{R_{n}} \frac{L_x}{W}=\frac{\Phi_0 
B_{\rm c2}}{
R_{n}}  \sim 10^{-18} \ \mathrm{kg/s}
,
\end{equation}
with $R_{n, {\rm sheet}} \simeq 300\ \mathrm{\Omega}$~\cite{Saito}.

The parameter $U_0$, the height of the pinning potential of the superconducting vortex, is 
estimated by the plateau of $R
$-$B$ curve in weak $B$ region. 
The 
plateau disappears  for $j_{\mathrm{sheet}} > j_{\rm c} \simeq 3 
\mathrm{A/m}$ \cite{Saito_private}, where $j_{\rm c}$ is the critical current density for vortex depinning.
Depinnig transition occurs when the pinning potential is well tilted by the external force, namely; 
\begin{equation}
U_0 \sim FL=j_{\mathrm{sheet}} \Phi_0 L \simeq 4\ \mathrm{meV}
.
\end{equation}
Another estimation of $U_0$ is from the thermodynamic upper critical field. The vortex energy 
per unit volume is $B_{c2}^2/(2\mu_0)$. 
When the vortex has the overlap with normal core in 
the pinning center, the energy reduces. By using $B_{c2} \simeq 0.1\ \mathrm{T} $~\cite{Saito}, 
we obtain
\begin{equation}
U_0\sim \frac{B_{c2}^2}{2\mu_0} \pi \xi ^2 c_z \simeq 5 \ \mathrm{meV}
,
\end{equation}
with $\xi\simeq 8 \mathrm{nm}$ and $c_z\simeq 1 \mathrm{nm}$ being the in-plane coherence 
length (normal core radius) and the lattice constant in the thickness direction, respectively. 
These two estimations are 
very consistent. We note $U_0\simeq 5 \ \mathrm{meV}\simeq 58  \ \mathrm{K}$ is much larger 
than the transition temperature therefore low temperature limit, $\beta U_0 \gg 1$, is the 
realistic situations in MoS$_2$.

\begin{figure}[t]
\begin{center}
\includegraphics[width=70mm]{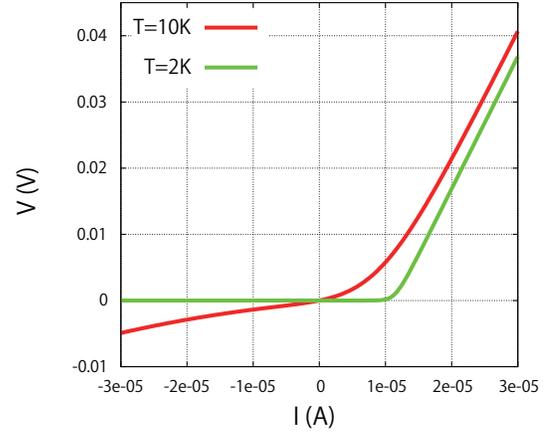}
\caption{
$I$-$V$ characteristic curve 
for the monolayer MoS$_2$ at $T=10$K and $T=2$K.
}
\label{fig:IVratchet}
\end{center}
\end{figure}

Calculated $I$-$V$ characteristic curve is shown in Fig.~\ref{fig:IVratchet}, which 
displays a 
strong rectification behavior even at a measurable temperature. For larger current such that 
$U_0<FL$, the expansion with respect to $F$ or equivalent to $I$ is no longer valid. In this 
regime, the $I$-$V$ characteristics are strongly non-linear and hence the higher harmonics 
becomes relevant. At the critical current where $U_0=F_cL$, the potential becomes a 
multi-step function and therefore the velocity begins to grow rapidly.

The temperature dependence of $\gm$-value at low $T$ is given by $\gm_{\rm S}'\propto 1/T$ for the ratchet mechanism according to Eq.~\eqref{eq:gamma_ratchet}
in the $U_0\gg T$ limit.
With the original expression in Eq.~\eqref{eq:gamma_ratchet},
the estimated $\gm$-values
for the sample size $L_xWc_z = 3{\rm \mu m}\times 3{\rm \mu m} \times 1{\rm n m}$ are given by
$\gm_{\rm S}'\simeq 8\times 10^5 {\rm A^{-1}}$ at $T=10$K and
$\gm_{\rm S}'\simeq 1\times 10^6 {\rm A^{-1}}$ at $T=2$K.
Although these magnitudes are much larger than the ones in the experimental observation \cite{Wakatsuki}, we can obtain the more close values by controlling the spatial asymmetry in the sawtooth potential.
Namely, by tuning the potential from the asymmetric case in Eq.~\eqref{eq:saw} to symmetric case continuously, the $\gm$-value is monotonically decreased down to zero.

At sufficiently low temperatures, on the other hand, the quantum nature of the vortices plays an important role.
In this case the wave character of vortices appears, which will modify the above physical picture.
This point remains to be explored in the future.

\section{Summary and conclusions} \label{Sec:Conclusion}

We have theoretically investigated the nonreciprocal charge transports in the two-dimensional superconductors without inversion symmetry.
We have taken the concrete examples such as the Rashba superconductors and topological insulator surface with in-plane magnetic fields, and the monolayer transition metal dichalcogenide (MoS$_2$) with out-plane magnetic field.
The nonreciprocal properties of superconductors are reflected 
in 
the $I$-$V$ characteristics with the form 
$V=a_1I+a_2I^2+a_3I^3+a_4I^4$, and the odd-order terms represent the nonreciprocal responses.
These coefficients
are clarified in the temperature range both above and below the mean-field transition temperature.
Table \ref{fig:table} summarizes our obtained results.

The nonreciprocal transport signals in the normal regions well above the mean-field transition temperature $T_0$ crossover into superconducting fluctuation contribution (paraconductivity).
We have newly investigated the 
topological insulator based systems which have cubic term in the GL free energy, in addition to the previously investigated 
Rashba superconductors and transition metal dichalcogenide.
The $\gm$-values for all the systems are much enhanced compared to the normal state, which is attributed to the appearance of the small energy scale $T_0$ ($\ll E_{\rm F}$).
The ratio of $\gm$-values between normal and superconducting fluctuation region is in general written as $(E_{\rm F}/T_0)^m$ with $m \geq 1$ being an integer depending on the system.

Below $T_0$, the amplitude of the superconducting order parameters is developed and only the phase degrees of freedom are left. 
Then the vortex dynamics plays an important role for $T\lesssim T_0$.
There are two different kinds of vortex behaviors.
First, for systems with in-plane magnetic fields, the vortices are thermally generated and are bound 
below
the Kosterlitz-Thouless transition temperature $T_{\rm KT}$.
Due to the inversion symmetry breaking in the system, the friction force and the number of vortices for $T_{\rm KT}<T<T_0$ are different depending on the direction of external uniform current, to produce the nonreciprocal charge transport.
These effects have effectively been described by the renormalization of the superfluid density.
The $\gm$-value, which is defined by $V = a_1 I (1+\gm B I)$ as in the fluctuation regime, is identified to have the temperature-dependent forms 
$\gm \propto (T_0 - T)^{-1}$ near $T_0$
and 
$\gm \propto (T-T_{\rm KT})^{-3/2}$ near $T_{\rm KT}$, which originate from the modified friction coefficient and the modified KT transition temperature, respectively.
Accordingly, we expect the two-peak structure of the temperature dependence of the $\gamma$-value in this case.

For the system with out-plane magnetic field, the number of vortices with a same vorticity is determined by the strength of the external field $B$ and the Kosterlitz-Thouless transition does not exist.
The nonreciprocal signal $\gm$ is now characterized by the $I$-$V$ relation with the form $V = a_1' BI (1+\gm B I)$.
The renormalization of superfluid density works also for this system, and we have derived the modified friction force for moving vortices and the corresponding $\gm$-value has the form $\gm \propto (T_0-T)^{-1}$.
This phenomenological approach is further justified by the time-dependent Ginzburg-Landau theory.
On the other hand, we have also investigated the other effect with ratchet potential for vortices.
Here the magnetic field plays a role only for creating the vortex, and then the $\gm$-value is characterized by the relation $V = a_1' BI (1+\gm_{\rm S}' I)$.
We have considered one dimensional motion driven by the external transport current in the sawtooth potential, and have found that the resultant signal $\gm'$ can have a comparable value with the experiments in MoS$_2$.

We have thus systematically clarified the characteristic transport properties for two-dimensional noncentrosymmetric superconductors.
The knowledge of this paper is useful for the further exploration of nonreciprocal phenomena in superconductors both theoretically and experimentally.

\section*{Acknowledgments}

We are grateful to Yu Saito, Toshiya Ideude, Yuki Itahashi, Yasuhiro Iwasa for fruitful discussions.
S.H. acknowledges Yusuke Kato for useful conversation on vortex dynamics.
R.W. was supported by the Grants-in-Aid for Japan Society for the Promotion of 
Science (JSPS) No. JP15J09045.
K.H. was supported by JSPS through a research fellowship for young scientists and the Program for Leading Graduate Schools (MERIT).
N.N. was supported by Ministry of Education, Culture, Sports, Science, and 
Technology Nos. JP24224009 and JP26103006, the Impulsing Paradigm Change through 
Disruptive Technologies Program of Council for Science, Technology and Innovation 
(Cabinet Office, Government of Japan), and Core Research for Evolutionary 
Science and Technology (CREST) No. JPMJCR16F1.

\appendix

\section{Derivation of the Ginzburg-Landau free energy}
Following Ref.~\onlinecite{BauerSigrist}, we derive the GL free energy for the 
model in Eq. (\ref{Eq:rashba}).
Especially, we focus on the case where the Fermi energy is on the conduction band.
The free energy is given by
\begin{widetext}
\begin{equation}
F = \int \frac{d^2 \bq}{\left(2\pi\right)^2} \left[
\frac{1}{g} - T \sum_{\omega_n} \int \frac{d^2\bk}{\left(2\pi\right)^2} 
G\left(\bk, i\omega_n\right) G\left(-\bk+\bq, -i\omega_n\right) \right] 
\left|\Psi_{\bq}\right|^2, \label{Eq:free}
\end{equation}
\end{widetext}
with $G\left(\bk, i\omega_n\right) = \left(i \omega_n - \xi_{\bk}\right)^{-1}$ being the 
Matsubara Green's function.
The product of the Green functions is simplified as
\begin{align}
& \int \frac{d^2\bk}{\left(2\pi\right)^2} G\left(\bk, i\omega_n\right) G\left(-\bk+\bq, 
-i\omega_n\right) \notag \\
& = \int \frac{d^2\bk}{\left(2\pi\right)^2} \frac{1}{i \omega_n - \xi_{\bk}} 
\frac{1}{-i \omega_n - \xi_{-\bk+\bq}} \notag \\
& = - \int \frac{d^2\bk}{\left(2\pi\right)^2}
\frac{1}{i\omega_n - \xi^0_{\bk} - \Omega_1\left(\bk\right)}
\frac{1}{i\omega_n + \xi^0_{\bk} + \Omega_2\left(\bk, \bq\right)},
\end{align}
where
\begin{align}
\xi^0_{\bk} &= \frac{\bk^2}{2m} + \alpha 
|\bm k|
- \EF, \\
\Omega_1\left(\bk\right) &= \xi_{\bk} - \xi^0_{\bk}, \\
\Omega_2\left(\bk, \bq\right) &= \xi_{-\bk+\bq} - \xi^0_{\bk}.
\end{align}
We first integrate by $\xi^0_{\bk}$
\begin{align}
& - \int \frac{d^2\bk}{\left(2\pi\right)^2}
\frac{1}{i\omega_n - \xi^0_{\bk} - \Omega_1\left(\bk\right)}
\frac{1}{i\omega_n + \xi^0_{\bk} + \Omega_2\left(\bk, \bq\right)} \notag \\
& \approx - \nu_1 \left< \int d\xi \frac{1}{i \omega_n - \xi - \Omega_1\left(\bk\right)} 
\frac{1}{i \omega_n + \xi + \Omega_2\left(\bk, \bq\right)} \right>_{\bk} \notag \\
& = \pi \nu_1 \left< \frac{1}{\left|\omega_n\right| + i \mathrm{sgn}
\left(\omega_n\right)\Omega\left(\bk, \bq\right)} \right>_{\bk},
\end{align}
where $\left<\cdots\right>_{\bk}$ is the momentum average over the Fermi surface.
When we consider the surface state of topological insulator, which we describe here for simplicity,
we only take the inner one of the two Fermi surfaces.
The Fermi wavenumber is $\kFa = -m\alpha + \sqrt{m\EFR}$, and the density of states is 
$\nu_1 = \frac{m}{2\pi}\left(1-\alpha\sqrt{m/\EFR}\right)$.
We have defined $\Omega\left(\bk, \bq\right) = \frac{1}{2}\left(\Omega_1
\left(\bk\right) - \Omega_2\left(\bk, \bq\right)\right)$.
Therefore, the second term in Eq. (\ref{Eq:free}) is
\begin{align}
& -\pi T \nu_1 \sum_{\omega_n} \left< \frac{1}{\left|\omega_n\right| 
+ i \mathrm{sgn}\left(\omega_n\right)\Omega\left(\bk, \bq\right)} \right>_{\bk} \notag \\
& \approx -\pi T \nu_1 \sum_{\omega_n} \left< \frac{1}{\left|\omega_n\right|} 
- \frac{\Omega\left(\bk, \bq\right)^2}{\left|\omega_n\right|^3} + 
\frac{\Omega\left(\bk, \bq\right)^4}{\left|\omega_n\right|^5} \right>_{\bk} \notag \\
&= -\nu_1 \left[ S_1\left(T\right) - S_3\left(T\right) \left<\Omega\left(\bk, \bq\right)^2\right> 
+ S_5\left(T\right) \left<\Omega\left(\bk, \bq\right)^4\right> \right],
\end{align}
with $S_k\left(T\right) = \pi T \sum_n \left|\omega_n\right|^{-k}$.
The function $S_k(T)$ is calculated as
\begin{align}
S_1\left(T\right) &= \log \frac{2e^{\gamma_\mathrm{E}}\Ec}{\pi T}, \\
S_3\left(T\right) &= \frac{7\zeta\left(3\right)}{4\left(\pi T\right)^2}, \\
S_5\left(T\right) &= \frac{31\zeta\left(5\right)}{16\left(\pi T\right)^4},
\end{align}
with $\gamma_\mathrm{E}$ being the Euler's constant and $\Ec$ being the cutoff energy.
Then, we calculate $\left<\Omega\left(\bk, \bq\right)^2\right>$ and 
$\left<\Omega\left(\bk, \bq\right)^4\right>$ up to $O\left(B_y \bq^4\right)$.
If we shift the momentum as $\bq \rightarrow \bq + \frac{2 B_y m}{|\bm k| + m \alpha}
\hat{\bm x}$,
we find
\begin{align}
\left<\Omega\left(\bk, \bq\right)^2\right> &= \frac{\left(\kFa+m\alpha\right)^2}{8m^2}
\bq^2 
\nonumber \\
&\ \ + \frac{3 \left(5\kFa + 3m\alpha\right)}{32m \kFa \left(\kFa+m\alpha\right)} 
B_y q_x \bq^2 + O\left(\bq^4\right), \\
\left<\Omega\left(\bk, \bq\right)^4\right> &= O\left(\bq^4\right).
\end{align}
This result is reasonable because the third order term in the momentum vanishes for 
$m \rightarrow \infty$.
The case of the valence band $\left(\EF < 0\right)$ can be obtained in a similar manner, and 
we can show that the free energy is obtained by replacing $\alpha$ with $-\alpha$, $B_y$ with 
$-B_y$, $\kFa = -m\alpha + \sqrt{m \EFR}$ with $\kFb = m\alpha-\sqrt{m \EFR}$ and $\nu_1 = 
\frac{m}{2\pi}\left(1-\alpha\sqrt{m/\EFR}\right)$ with $\nu_2 = 
\frac{m}{2\pi}\left(-1+\alpha\sqrt{m/\EFR}\right)$.
We can also derive the GL free energy for the model with the hexagonal warping [Eq. 
(\ref{Eq:hexagonal})] in the same way.

\section{Impurity effect in TMD}

The effect of superconducting fluctuation is more prominent for dirty samples according 
to the Ginzburg-Levanyuk criterion \cite{LarkinVarlamov}
\begin{align}
|\epsilon| \lesssim \left[ \frac{k_{\rm F}\xi_0}{(k_{\rm F}\xi)^D} \right]^{\frac{2}{4-D}}
,
\end{align}
where $\epsilon=(T-T_0)/T_0$ is the reduced temperature, $D$ is the dimension of the system.
$\xi$ is the coherence length for either clean or dirty samples. 
Here we consider the impurity effect on the GL equation with $\bm q$-cubic term originating from 
trigonal warping for MoS$_2$.
To deal with the impurities, we take quasi-classical Green function method \cite{Kopnin_book}. 
We introduce the normal and anomalous quasiclassical Green functions by 
$g(\hat {\bm k},\imu\omega_n;\bm r)$ and $f(\hat {\bm k}, \imu\omega_n; \bm r)$, 
respectively, where $\hat {\bm k}$ is the unit vector in the direction of $\bm k$ on the warped 
Fermi surface and $\omega_n=(2n+1)\pi T$ is the fermionic Matsubara frequency.
The gap equation is given by 
\begin{align}
\Delta (\bm r) = \pi \imu  \nu V^{\rm g}T \sum_n\la f(\hat {\bm k}, \imu\omega_n; \bm r) 
\ra_{{\bm k}}
, \label{eq:gap_eq}
\end{align}
where the bracket means the average with respect to ${\bm k}$.
$V^{\rm g}$ and $\nu$ are the attractive interaction parameter and density of states at the Fermi 
level, respectively.
To derive the GL theory, we expand the right-hand side of Eq.~\eqref{eq:gap_eq}. 
The terms without spatial derivatives are not affected by impurities, which is known as the Anderson 
theorem. Hence we only keep the linear term in $\Delta$ and consider spatial derivatives. With this 
condition we can use $g(\hat {\bm k},\imu\omega_n;\bm r)={\rm sgn\,} \omega_n$, since it 
does not have the linear term of $\Delta$. The anomalous Green function is described by the 
Eilenberger equation: 
\begin{align}
i \bm v(\hat {\bm k}) \cdot \bm \nabla f + 2 i\omega_n f - 2\Delta g+ \frac{i}{\tau} (\langle g 
\rangle_{{\bm k}} f - \langle f \rangle_{{\bm k}} g) = 0
,
\end{align}
where $\bm v$ is the Fermi velocity.
The self energy from impurities has been included by the self-consistent Born approximation.
The Zeeman energy can be accounted for by the simple replacement 
$i\omega_n \rightarrow i\omega_n + 
B_z$, and we do not write this explicitly 
for the moment.
Now we perform the gradient expansion as $f=f_0 + f_1 + f_2 + f_3 + \cdots$.
The zeroth and first-order terms can be explicitly written as
\begin{align}
\langle f_0 \rangle_{{\bm k}} &= f_0 = \frac{\Delta {\rm sgn\,}\omega_n}{i\omega_n}
, \\
\langle f_1 \rangle_{{\bm k}} &= - \frac{i \langle \bm v \rangle_{{\bm k}} \cdot \bm 
\nabla f_0}{2i\omega_n}
,
\end{align}
where $\tilde \omega_n = \omega_n + \frac{1}{2\tau} {\rm sgn\,}\omega_n$.
We conclude $\la f_1 \ra_{{\bm k}}=0$ since $\langle \bm v \rangle_{{\bm k}} =0$ 
is satisfied.
The higher-order terms can also be derived as
\begin{align}
\langle f_2 \rangle_{{\bm k}} &= \frac{\langle (i\bm v \cdot \bm \nabla)^2 f_0 
\rangle_{{\bm k}}}{4i\omega_n i\tilde \omega_n}
,\\
\langle f_3 \rangle_{{\bm k}} &= - \frac{\langle (i\bm v \cdot \bm \nabla)^3 f_0 
\rangle_{{\bm k}}}{4i\omega_n (i\tilde \omega_n)^2}
.
\end{align}
$\la f_3 \ra_{{\bm k}}$ can be finite if the system has a trigonal warping.
To be compatible with the results in the previous sections, there are the relations
\begin{align}
\langle (i\bm v \cdot \bm \nabla)^2 \rangle_{{\bm k}} &= C_2 \bm \nabla ^2
,\\
\langle (i\bm v \cdot \bm \nabla)^3 \rangle_{{\bm k}} &= \imu C_3 
\partial_x (\partial_x^2-3\partial_y^2)
.
\end{align}
The real constants $C_2$ and $C_3$ are determined to be consistent with the expressions in the 
clean limit which has already been obtained in Ref.~\onlinecite{Wakatsuki}.

We now replace $i\omega_n$ by $i\omega_n + 
B_z$ to include the Zeeman energy, 
and take
the lowest order contribution of the external magnetic field $B_z$.
We substitute these expressions into the gap equation.
The coefficient $\gm(\tau)$ of $\bm q$-square term and $K(\tau)$ of $\bm q$-cubic term in the 
GL equation, which are now dependent on the mean-free time $\tau$, are given by
\begin{align}
&\gm(\tau)/\gm(\infty) = \sum_n \frac{1}{|\omega_n|^2 (|\omega_n| + \frac{1}{2\tau})}
\left/
\sum_n \frac{1}{|\omega_n|^3}
\right.
, \\
&K(\tau)/K(\infty) =
\nonumber \\
&\frac 1 2 \sum_n \frac{1}{|\omega_n|^2 (|\omega_n| + \frac{1}{2\tau})^2} 
\left( \frac{1}{|\omega_n|} + \frac{1}{|\omega_n| + \frac{1}{2\tau}} \right)
\left/
\sum_n \frac{1}{|\omega_n|^5}
\right.
,
\end{align}
to obtain the results with impurities.
These factors can be used at any purity of samples and go to unity for the clean limit.
The square coefficient here is not a new result which can be seen in e.g. Ref.~\cite{Kim_book}, 
and the cubic coefficient is first derived.
Particularly for the $\gamma$-value, the ratio between dirty 
($\tau T_0 \ll 1$) and clean 
($\tau T_0 \gg 1$) limits is given by
\begin{align}
\frac{\gamma^{\rm f}_{\rm S,dirty}}{\gamma^{\rm f}_{\rm S, clean}}
\sim \tau T_0
,
\end{align}
for superconducting fluctuation contribution.
For vortex flow contribution, on the other hand, the transport coefficients are dependent on 
$\sg_n$ ($\simeq n_e e^2 \tau/m$) and cannot be written in a simple way, but can be in 
general estimated from the above information.

\section{Effect of Landau level for paraconductivity in TMD}

\subsection{Formulation}

We here extend the calculation for the paraconductivity in the low-field limit to the case in the 
presence of quartic term and magnetic field.
The paraconductivity from superconducting fluctuations can be evaluated at arbitrary strength of the 
magnetic field by considering the Landau levels.
Let us begin with the GL free energy
\begin{align}
F &= \int \diff \bm r 
\Psi^* \left[
a + \frac{\bm P^2}{2m^*} + \Lambda B_z (P_x^3 - 3\overline{P_xP_y^2})
\right] \Psi
,
\end{align}
where $\bm P = -\imu \bm \nabla - e^*\bm A$ with $e^*=2e$ and $m^*=2m$.
The quartic term with $|\Psi|^4$ can be effectively included in the square term by using the 
self-consistent harmonic approximation and is dropped here.
We will explain this point later.
The spatially averaged supercurrent is given in the simple form
\begin{align}
J_{sx} 
&= - |e^*|\int \frac{\diff \bm r}{\Omega} \Psi^* \left[ \frac{1}{m^*} P_x + 3\Lambda B_z
(P_x^2-P_y^2)  \right] \Psi
,
\end{align}
where $\Omega = \int \diff\bm r 1$ is a two-dimensional system volume.
The $y$-component can be constructed by symmetry and we do not consider here.
We choose the vector potential as $A_x = -E_x t$ and $A_y = -E_y t + B_zx$.
Now we expand the complex function $\Psi$ as
\begin{align}
\Psi(\bm r, t) &= \sum_{kn} c_{kn}(t) \epn^{-\imu ky}\epn^{\imu |e|E_x t[x - x_{0k}(t/2)]} 
h_n(x-x_{0k}(t))
,\\
x_{0k}(t) &= \frac{k+|e^*|E_y t}{|e^*|B_z}
,
\end{align}
where $h_n(x)$ is an eigenfunction of the one-dimensional harmonic oscillator with the quantum number $n$.
We can show the following relations
\begin{align}
P_x \Psi &= -\imu \sqrt{\frac{m^*\omega}{2}} (b-b^\dg) \Psi
, \\
\frac{\bm P^2}{2m}\Psi &=  \omega (b^\dg b + \tfrac 1 2) \Psi
, \\
(P_x^3 - 3\overline{P_xP_y^2})\Psi &= 4\imu \left( \frac{m^* \omega}{2} \right)^{3/2} 
(b^3 - b^{\dg 3}) \Psi
, \\
(P_x^2 - P_y^2)\Psi &= - m^* \omega (b^2 + b^{\dg 2}) \Psi
,
\end{align}
with $\omega = |e^*|B_z/m^*$ being a cyclotron frequency.
We have used the relations
\begin{align}
\partial_{x} h_n(x) &= \sqrt{\frac{m^*\omega}{2}} (b-b^\dg) h_n (x)
, \\
x h_n(x) &= \sqrt{\frac{1}{2m^*\omega}} (b+b^\dg) h_n(x)
,
\end{align}
The operators $b$ and $b^\dg$ act only on $h_n$ as 
\begin{align}
b h_n &= \sqrt{n} h_{n-1} 
,\\
b^\dg h_n &= \sqrt{n+1} h_{n+1}
.
\end{align}
Let us consider the TDGL equation in the presence of thermal fluctuations:
\begin{align}
-\Gamma \partial_t \Psi = \frac{\delta F}{\delta \Psi^*} - f
, \label{eq:TDGL_LL}
\end{align}
We note that $\Gamma$ here is different from the one in Eq.~\eqref{eq:TDGL_vtx} by a constant factor.
Equation \eqref{eq:TDGL_LL} can be rewritten in terms of $c_{kn}(t)$ as
\begin{align}
-\Gamma\partial_t c_n
&= [a+\omega(n+\tfrac 1 2 )] c_n
- f'_n
\nonumber \\
&\ \ 
+ 4\imu \Lambda B_z \left(\frac{m^*\omega}{2}\right)^{3/2}
\left[
\sqrt{\permutation{n+3}{3}} c_{n+3} - \sqrt{\permutation{n}{3}} c_{n-3}
\right]
\nonumber \\
&\ \ 
+ \frac{\imu |e|\Gamma}{\sqrt{2m^*\omega}}\left[
\sqrt{n+1} E c_{n+1} + \sqrt{n} E^* c_{n-1}
\right] 
,
\end{align}
where we have defined the complex electric field $E=E_x+\imu E_y$ and have omitted $k$.
The symbol $\permutation{n}{m}=n!/(n-m)!$ is the permutation.
This equation can be solved perturbatively.
We expand the solution as
\begin{align}
c_n(t) &= \sum_{p,q=0}^{\infty} c_n^{pq}(t)
,
\end{align}
where $ c_n^{pq}$ is a contribution of $O(E^p\Lambda^q)$, and each term satisfies the 
following recursive equation:
\begin{align}
&c_n^{pq}(t) = \frac{\imu}{\Gamma} \int_{-\infty}^t d t' \left[
\beta (\sqrt{\permutation{n+3}{3}}c_{n+3}^{p,q-1}-\sqrt{\permutation{n}{3}}c_{n-3}^{p,q-1})
\right.
\nonumber \\
& \left.
+ \alpha (\sqrt{n+1}Ec_{n+1}^{p-1,q}+\sqrt{n}E^*c_{n-1}^{p-1,q})
\right]\epn^{(t'-t)A_n}
. \label{eq:langevin}
\end{align}
Here we have defined
\begin{align}
&\al = - \frac{|e|\Gamma}{\sqrt{2m^*\omega}}
,\ \ 
\beta = - 4\Lambda B_z \left(\frac{m^*\omega}{2}\right)^{3/2}
, \\
&A_n = \frac{a + \omega(n+\tfrac 1 2)}{\Gamma}
,
\end{align}
to make the notation simple.

We rewrite the current along $x$-direction as
\begin{align}
J_{sx} &= \frac{|e^*|^2 B_z}{\pi} \sum_{n=0}^{E_{\rm c}/\omega}  \left[
\sqrt{\frac{\omega}{2m^*}} \sqrt{n+1} \  \imag \la c^*_{n+1} c_n \ra
\right.
\nonumber \\
&\ \ \left.
+ 3 \Lambda B_z m^*\omega \sqrt{\permutation{n+2}{2}} \  \real \la c^*_{n+2} c_n \ra
\right]
,
\end{align}
where we have used the relation for the number of degeneracy:
\begin{align}
\sum_k 1 &= \frac{\Omega}{2\pi\ell_B^2}
,
\end{align}
where $\ell_B = \sqrt{1/|e^*|B_z}$ is the magnetic length.
The cutoff energy is given by $E_{\rm c} \sim T_0$.

\subsection{Evaluation of current}
We first consider the zeroth-order component:
\begin{align}
\la c_{n}^{00*} (t_1) c_{n}^{00} (t_2) \ra &= \frac{ T/\Gamma}{A_n} \epn^{-|t_1-t_2|A_n}
. \label{eq:zeroth}
\end{align}
To evaluate the current we need only the equal-time component of the forms $\la c^*_{n+1} c_n 
\ra$ and $\la c^*_{n+2} c_n \ra$.
The $O(E^1)$ components are calculated from Eqs.~\eqref{eq:langevin} and \eqref{eq:zeroth} as
\begin{align}
\la c_{n+1}^{10*} c_{n}^{00} \ra
&= - \frac{\imu \al E  T}{\Gamma^2} \sqrt{n+1} \frac{1}{A_n(A_n + A_{n+1})}
, \\
\la c_{n-1}^{10*} c_{n}^{00} \ra
&= - \frac{\imu \al E^*  T}{\Gamma^2} \sqrt{n} \frac{1}{A_n(A_n + A_{n-1})}
.
\end{align}
The $O(E^2)$ components are calculated as
\begin{align}
\la c^{10*}_{n+2} c^{10}_{n} \ra
&= \frac{\al^2E^2 T}{\Gamma^3}\sqrt{\permutation{n+2}{2}} \  H_1(n+1, n+2, n)
, \\
\la c^{00*}_{n+2} c^{20}_n \ra
&= \frac{-\al^2E^2 T}{\Gamma^3} \sqrt{\permutation{n+2}{2}} \  H_2 (n+2, n, n+1)
, \\
\la c^{00*}_{n-2} c^{20}_n \ra
&= \frac{-\al^2 (E^*)^2 T}{\Gamma^3} \sqrt{\permutation{n}{2}} \  H_2 (n-2, n, n-1)
,
\end{align}
where 
\begin{align}
H_1(i,j,k) &= 
\frac{1}{A_i(A_j+A_k)}\left[
\frac{1}{A_i+A_k} 
+ \frac{1}{A_i+A_j}
\right] 
, \\
H_2(i,j,k) &= 
\frac{1}{A_i(A_i+A_k)(A_i+A_j)}
.
\end{align}
The $O(E^2\Lambda^1)$ components are also calculated as
\begin{widetext}
\begin{align}
\la c^{11*}_{n+1} c^{10}_n \ra
=&
- (C^{21})^*
\left[
\frac{\permutation{n+1}{3}}{\sqrt{n+1}}  F_1(n-1, n+1, n, n-2)
+ 
\frac{\permutation{n+2}{3}}{\sqrt{n+1}}  F_1(n-1, n+1, n, n+2)
\right]
, \\
\la c^{11*}_{n-1} c^{10}_n \ra
=&
-C^{21}
\left[
\frac{\permutation{n+2}{3}}{\sqrt n}  F_1(n+1, n-1, n, n+2)
+ 
\frac{\permutation{n+1}{3}}{\sqrt n}  F_1(n+1, n-1, n, n-2)
\right]
, \\
\la c^{20*}_{n+1} c^{01}_n \ra
=&
(C^{21})^*
\frac{\permutation{n+3}{3}}{\sqrt{n+1}}  F_1(n+3, n+1, n, n+2)
, \\
\la c^{20*}_{n-1} c^{01}_n \ra
=&
C^{21}
\frac{\permutation{n}{3}}{\sqrt n}  F_1(n-3, n-1, n, n-2)
, \\
\la c^{21*}_{n+1} c^{00}_n \ra
=& 
(C^{21})^*
\left[
\frac{\permutation{n+1}{3}}{\sqrt{n+1}}  F_2(n, n+1, n-2, n-1)
+ \frac{\permutation{n+3}{3}}{\sqrt{n+1}}  F_2(n, n+1, n+2, n+3)
+ \frac{\permutation{n+2}{3}}{\sqrt{n+1}}  F_2(n, n+1, n+2, n-1)
\right]
, \\
\la c^{21*}_{n-1} c^{00}_n \ra
=& 
C^{21}
\left[
\frac{\permutation{n+2}{3}}{\sqrt n}  F_2(n, n-1, n+2, n+1)
\frac{\permutation{n}{3}}{\sqrt n}  F_2(n, n-1, n-2, n-3) 
\frac{\permutation{n+1}{3}}{\sqrt n}  F_2(n, n-1, n-2, n+1)
\right]
, 
\end{align}
where we have defined the complex constant by $C^{21} = \imu \al^2\beta E^2T/\Gamma^4$ 
and the functions $F_{1,2}$ by
\begin{align}
F_1(i,j,k,l) &= 
\frac{1}{A_i(A_j+A_k)}\left[
\frac{1}{(A_i+A_l)(A_k+A_l)}
+ \frac{1}{(A_i+A_k)(A_k+A_l)}
+ \frac{1}{(A_i+A_l)(A_i+A_j)}
\right]
, \\
F_2(i,j,k,l) &= 
\frac{1}{A_i(A_i+A_l)(A_i+A_k)(A_i+A_j)}
.
\end{align}
The quantities other than the listed above can be evaluated using a complex conjugation relations.
Substituting the above expressions into the current, we obtain the linear and nonlinear paraconductivities.
We define the transport coefficients by 
$J_{sx} = \sigma_1 E_x + \sigma_2 (E_x^2 - E_y^2) + O(E^3)$ and
each coefficient is given  by
\begin{align}
\sigma_1 &= \frac{|e^*|^2  \omega  T}{2\pi\Gamma} \sum_n \frac{n+1}{A_n+A_{n+1}} 
\left( \frac{1}{A_n} - \frac{1}{A_{n+1}} \right)
, \\
\sigma_2 &= - \frac{|e^*|^3m^*\Lambda B_z  \omega  T}{2\pi\Gamma^2} \sum_{n} 
[ \omega X(n) + 3 \Gamma Y(n)]
, 
\end{align}
where
\begin{align}
X(n)&=
 \permutation{n+1}{3} F_1(n-1, n+1, n, n-2)
+\permutation{n+2}{3} F_1(n-1, n+1, n, n+2)
\nonumber \\ 
&
+\permutation{n+3}{3} F_1(n+2, n, n+1, n+3)
+\permutation{n+2}{3} F_1(n+2 ,n, n+1, n-1)
\nonumber \\ 
&
-\permutation{n+3}{3} F_1(n+3, n+1, n, n+2)
-\permutation{n+1}{3} F_1(n-2, n, n+1, n-1)
\nonumber \\ 
&
-\permutation{n+1}{3} F_2(n, n+1, n-2, n-1)
-\permutation{n+3}{3} F_2(n, n+1, n+2, n+3)
-\permutation{n+2}{3} F_2(n, n+1, n+2, n-1)
\nonumber \\ 
&
-\permutation{n+3}{3} F_2(n+1, n, n+3, n+2)
-\permutation{n+1}{3} F_2(n+1, n, n-1, n-2)
-\permutation{n+2}{3} F_2(n+1, n, n-1, n+2)
, \\
Y(n) &= \permutation{n+2}{2} [ H_2(n+2,n,n+1) + H_2(n,n+2,n+1) - H_1(n+1,n+2,n) ]
.
\end{align}
\end{widetext}
One can check that the function in the $n$-summation behaves as $O(n^{-2})$ or faster at large $n$
, so we do not need the cutoff for convergence.
The ordinary paraconductivity $\sg_1$ here reproduces the results derived in the previous work 
\cite{Abrahams71}.

\subsection{Effect of quartic term}

Here we consider the quartic term in GL free energy
\begin{align}
F_4 &= \frac b 2  \int \diff \bm r \,
|\Psi|^4.
\end{align}
We use the self-consistent harmonic approximation \cite{Chaikin_book}:
\begin{align}
|\Psi|^4 = \Psi^* \Psi^* \Psi \Psi \simeq 2\la |\Psi|^2 \ra |\Psi|^2
.
\end{align}
Hence the coefficient $a$ is replaced by $a' = a + b\la |\Psi|^2 \ra$.
We have the self-consistent equation to determine $a'$:
\begin{align}
\la |\Psi|^2 \ra &= \frac{1}{\Omega} \sum_{kn} \la |c_{kn}|^2 \ra
= \frac{m^* T}{2\pi} \sum_{n=0}^{E_{\rm c}/ \omega} \frac{1}{n + \frac 1 2 
+ \frac{a + b\la |\Psi|^2 \ra}{\omega}}
.
\end{align}
Note that here we need the cutoff energy $E_{\rm c}\sim  T_0$ for convergence, 
and have neglected the $\bm P$-cubic term which is irrelevant in the leading-order contribution for the equilibrium case.
With this consideration, the finite transition temperature in mean-field theory is washed away in the 
two-dimensional system reflecting the Mermin-Wagner theorem.

\begin{figure}[t]
\begin{center}
\includegraphics[width=80mm]{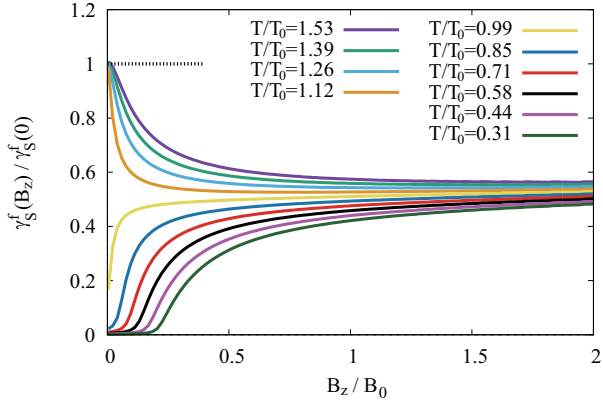}
\caption{
Magnetic field $B_z$ dependence of the $\gm$-value from paraconductivity.
The unit for the magnetic field is $B_0 = m^*T_0/ |e^*|$.
The parameters are chosen as $a(T)=0.02(T-T_0)$, $bm^* = 0.005$, and $E_{\rm c} = 4 T_0$.
}
\label{fig:B_depend}
\end{center}
\end{figure}

Figure~\ref{fig:B_depend} shows the exemplary results for the magnetic field dependence of 
$\gm_{\rm S}(B_z) = \sg_2/\sg_1^2 B_z W$ which is normalized by the value at $B_z=0$.
Above the mean-field transition temperature $T_0$, the $\gm$-value decreases with increasing $B_z$, 
and it becomes nearly half at high fields.
On the contrary, below $T_0$, the $\gm$-value is small at low $B_z$ and increases by applying 
magnetic field.
Although $\gm_{\rm S}(B_z)/\gm_{\rm S}(0)$ goes to 1 at zero field, the corresponding $B_z$-range 
is so narrow 
that such regime (i.e. $a'\gg \omega$) cannot be seen practically.
This is because the value of $a'$ for $T< T_0$ remains positive but is so tiny, and therefore the 
behavior is very sensitive to the small value of $B_z$.
We note that in actual systems the component of normal conductivity is also finite and modifies the 
$\gm$-value down to zero at high magnetic fields.

\end{document}